\documentclass[twocolumn]{aastex63}
\usepackage{enumitem}
\setlist[itemize]{leftmargin=4ex, topsep=2pt, itemsep=2pt, parsep=0pt}
\setlist[enumerate]{leftmargin=4ex, topsep=2pt, itemsep=2pt, parsep=0pt}

\usepackage{graphicx} 
\usepackage{natbib} 
\usepackage{amsmath} 
\usepackage{multirow}
\bibpunct{(}{)}{;}{a}{}{,}

\newcommand{\ee}[1]{\mbox{${} \times 10^{#1}$}}
\newcommand{\eten}[1]{\mbox{$10^{#1}$}}

\newcommand{\blank}{\mbox{ }}  
\newcommand{\mean}[1]{\mbox{$\langle#1\rangle$}} 
\newcommand{\jj}[2]{\mbox{$J = #1\rightarrow#2$}}
\newcommand{\tastar}{\mbox{$T_{\rm A}^{*}$}}
\newcommand{\tmb}{\mbox{$T_{\rm mb}$}}
\newcommand{\msun}{\mbox{M$_\odot$}}
\newcommand{\msunmyr}{\mbox{M$_\odot$} Myr$^{-1}$}

\newcommand{\sigmadense}{\mbox{$\Sigma_{\rm dense}$}}
\newcommand{\nbar}{\mbox{$\bar{n}$}}

\newcommand{\coo}{\mbox{$^{13}$CO}}

\newcommand{\kms}{km s$^{-1}$}
\newcommand{\kkms}{K km s$^{-1}$}

\newcommand{\hh}{\mbox{{\rm H}$_2$}}
\newcommand{\hcop}{\mbox{{\rm HCO}$^+$}}
\newcommand{\form}{\mbox{{\rm H$_2$CO}}}
\newcommand{\hcopi}{\mbox{{\rm H$^{13}$CO}$^+$}}
\newcommand{\hcni}{\mbox{{\rm H$^{13}$CN}}}
\newcommand{\nthp}{\mbox{{\rm N}$_2${\rm H}$^+$}}
\newcommand{\hii}{\mbox{\ion{H}{2}}}

\newcommand{\mcloud}{\mbox{$M_{\rm cloud}$}}

\newcommand{\mdense}{\mbox{$M_{\rm dense}$}}
\newcommand{\mdv}{\mbox{$M_{\rm dv}$}}
\newcommand{\rdense}{\mbox{$r_{\rm dense}$}}

\newcommand{\av}{\mbox{$A_{\rm V}$}}
\newcommand\cmv{\mbox{cm$^{-3}$}}
\newcommand\cmc{\mbox{cm$^{-2}$}}

\newcommand{\tk}{\mbox{$T_{\rm K}$}}

\newcommand{\msunpc}{\mbox{M$_\odot$ pc$^{-2}$}}
\newcommand{\ldense}{\mbox{$L_{\rm dense}$}}
\newcommand{\ltot}{\mbox{$L_{\rm tot}$}}
\newcommand{\lhcn}{\mbox{$L({\rm HCN})$}}
\newcommand{\lhcop}{\mbox{$L({\rm \hcop})$}}
\newcommand{\ain}{\mbox{$\alpha_{\rm in}$}}
\newcommand{\at}{\mbox{$\alpha_{\rm tot}$}}
\newcommand{\avd}{\mbox{$\alpha_{\rm dv}$}}
\newcommand{\fl}{\mbox{$f_{\rm L}$}}
\newcommand{\lunit}{\mbox{K \kms\ pc$^{2}$}}
\newcommand{\clouda}{\mbox{G034.158$+$00.147}}
\newcommand{\cloudb}{\mbox{G034.997$+$00.330}}
\newcommand{\cloudc}{\mbox{G036.459$-$00.183}}
\newcommand{\cloudd}{\mbox{G037.677$+$00.155}}
\newcommand{\cloude}{\mbox{G045.825$-$00.291}}
\newcommand{\cloudf}{\mbox{G046.495$-$00.241}}
\newcommand{\mm}{\mbox{millimeter-wave}}

\shorttitle {Dense Gas}
\shortauthors{Evans et al.}

\begin{document}

\setcounter{table}{0}

\title{Star Formation Occurs in Dense Gas, but What Does ``Dense" Mean?}

\correspondingauthor{Neal J. Evans II}
\email{nje@astro.as.utexas.edu}
\author[0000-0001-5175-1777]{Neal J. Evans II}
\affil{Department of Astronomy 
The University of Texas at Austin 
2515 Speedway, Stop C1400
Austin, Texas 78712-1205, USA}
\affiliation{Korea Astronomy and Space Science Institute
 776 Daedeokdae-ro, Yuseong-gu 
 Daejeon, 34055, Korea}
\affiliation{Humanitas College, Global Campus, Kyung Hee University, Yongin-shi 17104, Korea}
\author{Kee-Tae Kim}
\affiliation{Korea Astronomy and Space Science Institute
 776 Daedeokdae-ro, Yuseong-gu 
 Daejeon, 34055, Korea}
\affiliation{University of Science and Technology, Korea (UST), 217 Gajeong-ro, Yuseong-gu, Daejeon 34113, Republic of Korea} 
\author{Jingwen Wu}
\affiliation{National Astronomical Observatories, Chinese Academy of Sciences, 20A Datun Road, Chaoyang District, Beijing, 100012, China}
\author{Zhang Chao}
\affiliation{National Astronomical Observatories, Chinese Academy of Sciences, 20A Datun Road, Chaoyang District, Beijing, 100012, China}
\author{Mark Heyer}
\affiliation{Department of Astronomy 
 University of Massachusetts  
Amherst, Massachusetts 01003, USA}
\author{Tie Liu}
\affiliation{Shanghai Astronomical Observatory, Chinese Academy of Sciences, 80 Nandan Road, Shanghai, 200030, PR China} 
\affiliation{Korea Astronomy and Space Science Institute
 776 Daedeokdae-ro, Yuseong-gu 
 Daejeon, 34055, Korea} 
\author{Quang Nguyen-Lu'o'ng}
\affiliation{McMaster University, 1 James St N, Hamilton, ON, L8P 1A2, Canada}
\affiliation{Graduate School of Natural Sciences, Nagoya City University, Mizuho-ku, Nagoya, Aichi 467-8601, Japan}
\affiliation{IBM Canada, 120 Bloor Street East, Toronto, ON, M4Y 1B7, Canada} 
\author{Jens Kauffmann}
\affiliation{Haystack Observatory MIT 
 99 Milstone Rd. 
 Westford, MA 01886 USA}

\begin{abstract}
We report results of a project
to map HCN and \hcop\  \jj10\ emission toward a sample of molecular clouds in
the inner Galaxy, all containing
dense clumps that are actively engaged in star formation.
We compare these two molecular line tracers with millimeter continuum 
emission and extinction, as inferred from \coo, as tracers of dense gas in molecular clouds.  
The fraction of the line luminosity from each tracer that comes from the
dense gas, as measured by $\av > 8$ mag, varies substantially 
from cloud to cloud.  In all
cases, a substantial fraction (in most cases, the majority)
of the total luminosity arises in gas below
the $\av > 8$ mag threshold and outside the region of strong mm continuum emission.
Measurements of \lhcn\ toward other galaxies will likely be dominated
by such gas at lower surface density. Substantial, even dominant, contributions
to the total line luminosity can arise in gas with densities typical of 
the cloud as a whole ($n \sim 100$ \cmv).
Defining the dense clump from the
HCN or  \hcop\ emission itself,  similarly to previous studies, leads to a wide range of clump properties, with some
being considerably larger and less dense than in previous studies.
\deleted{Rough estimates of the virial parameter confirm visual impressions that}
HCN and \hcop\ have similar ability to trace dense gas for the clouds in this sample. For the two clouds with low virial parameters, the \coo\ is definitely
a worse tracer of the dense gas, but for the other four, it is equally good
(or bad) at tracing dense gas.
\end{abstract}

\section{Introduction}

In pioneering work,
\citet{Gao:2004}
showed that the far-infrared luminosities in starburst galaxies followed a very
tight, linear correlation with the luminosities of HCN line emission.
\citet{Wu:2005}
showed that this relationship extended to massive, dense clumps in the
Milky Way, arguing that the fundamental unit of massive, clustered
star formation is such a massive, dense clump.
Subsequent studies have defined a ``threshold" surface density of
$\av > 8$ mag
(about 120 \msunpc) in nearby clouds
(\citealt{Heiderman:2010,Lada:2010,Lada:2012})
above which the vast majority of dense cores and YSOs are found.
\citet{Evans:2014}
compared various models of star formation to observations of the
nearby clouds and found that the mass of dense gas was the best
predictor of the star formation rate.
Most recently,
\citet{2016ApJ...831...73V}
showed that a similar result applied to more distant and massive
clouds in the Galactic Plane, using millimeter continuum emission
from the BGPS survey
\citep{Ginsburg:2013}
to measure the mass of dense gas. 
\citet{2016ApJ...831...73V}
found a substantial dispersion in the star formation rate per mass of dense
gas (0.50 dex), but the logarithmic averages of the 
star formation rate per mass of dense gas were in general agreement
for nearby clouds, inner Galaxy clouds, and extragalactic clouds.
The dispersion among the averages for all those was only 0.19 dex
(figure 11 of 
\citealt{2016ApJ...831...73V}), 
considerably lower than that for
total molecular gas probed by CO or \coo, 0.42 dex
(figure 12 of \citealt{2016ApJ...831...73V}).
The star formation rate per unit mass
of dense gas is thus remarkably constant over a huge range of scales and
conditions.
A recent detailed study of HCN emission toward other galaxies
\citep{2019ApJ...880..127J} 
also found a small dispersion (0.22 dex) in the star formation rate per mass of
dense gas. After accounting for galaxy-to-galaxy variations, the intra-galaxy
dispersion was only 0.12 dex. A smaller dispersion for measurements over
a whole galaxy is expected if the dispersion among clouds within a galaxy
is due to variations in evolutionary state of the molecular cloud
(e.g., \citealt{2018MNRAS.479.1866K}).

While the tight connection between dense parts of molecular clouds
and star formation is clear, we must better define what we mean by
``dense." The most direct measure exists for the nearby clouds, where
surface density can be determined by extinction maps of background stars.
For the Galactic Plane clouds, continuum emission by dust or line
emission by HCN was used. For other galaxies, HCN emission has been
the only tracer of dense gas in general use, although \hcop\
emission has also been explored
\citep{2011ApJS..196...12B,2015ApJ...814...39P,2019ApJ...880..127J}. 
The extinction maps are strictly sensitive to surface density, while
the dust continuum emission is sensitive also to temperature, and the molecular line emission, in addition to surface density and temperature, is 
sensitive to volume density and abundances. Temperature and molecular
abundances can depend on the radiation environment
\citep{2017A&A...604A..74S, 2017A&A...599A..98P}.
A detailed comparison of these various tracers of ``dense" gas can
clarify the situation.
In this paper, we will compare maps of HCN, \hcop, and \coo\ \jj10\ emission to the regions selected as dense by extinction or \mm\ continuum emission in a sample of Galactic Plane clouds. We refer to all the molecular lines as ``line tracers"
and the HCN and \hcop\ lines as ``dense line tracers" for convenience, while
noting that our purpose is to test the proposition that they trace gas of
the density relevant to star formation.

Specifically, we can learn what fraction of the
luminosity from the line tracers arises from low-level, extended emission 
from less dense gas.
\citet{2016ApJ...824...29S}
have  argued that most of the Galaxy's luminosity
of HCN arises from distributed,  very sub-thermal, emission rather than from
dense gas. Since observations of other galaxies would integrate large
areas of low-level emission, their HCN luminosity could be dominated
by the same gas that is probed by CO, complicating the connection found by
\citet{Wu:2005}
between dense clumps in the Milky Way and other galaxies.
Pioneering work by
\citet{1997ApJ...478..233H} 
showed that HCN emission is very weak compared to CO
($I_{\rm HCN}/I_{\rm CO} = 0.014\pm 0.020$) averaged over random
observations of the Galactic Plane. They further showed that maps
of HCN \jj10\ in nearby large molecular clouds were tiny in comparison
to the CO maps. This work argues against the idea of 
\citet{2016ApJ...824...29S}
but improved instrumentation now allows a much stronger test.
Recent work has shown that HCN emission can indeed arise from more
extended regions \citep{2017A&A...599A..98P, 2017A&A...605L...5K,
2017A&A...604A..74S}, but those studies were all toward clouds in 
the solar neighborhood (out to the distance of the Orion clouds).
\added{A recent study extends these results to the M17 cloud
\cite{2020ApJ...891...66N}.}

Maps of the full extent of \coo\ emission in inner Galaxy clouds
allow us to test directly the contribution of more diffuse
molcular gas to the HCN luminosity in a very different environment. The simultaneous observations of \hcop\ provide a direct comparison
of these two tracers. One might predict that \hcop\ \jj10\
traces more wide-spread gas of
lower mean density than does HCN \jj10\ 
because the critical  density for \hcop\ \jj10\ at $\tk = 20$ K
($n_{\rm cr} \approx 4.5\ee4$ \cmv)
is nearly an order of magnitude less than that for HCN \jj10\
($n_{\rm cr} \approx 3.0\ee5$ \cmv)
(\citealt{1999ARA&A..37..311E,2015PASP..127..299S}).
Comparison of these two tracers in a well-defined sample with
star formation rates will be useful for evaluating relations seen
in extragalactic studies of dense gas relations.
Studies of HCN and \hcop\ have found some evidence of
environmental effects in other galaxies, such as the
presence of an AGN
(see, e.g., \citealt{2015ApJ...814...39P} for a  discussion
of this issue.) 
It is important first to understand the relation between these
two tracers in more controlled environments.
We also evaluate whether \coo\ \jj10\ can trace the relevant gas for
star formation; while it has a much lower critical density
($n_{\rm cr} \approx 4.8\ee2$ \cmv), it is
easier to observe.

With spectrally resolved maps, we can also assess the balance between
gravity and turbulence, most simplistically captured in the virial
parameter. One attractive explanation for the low star formation efficiency
in molecular clouds is that most clouds are not gravitationally bound, 
but only relatively dense regions within them are bound 
\citep{2011MNRAS.413.2935D,2016ApJ...831...67B}.
\added{While studies differ, even the most massive clouds defined by
CO emission may be unbound \cite{2016ApJ...833...23N}.}
By calculating the virial ratio for the structures traced by the
different species, we may be able to shed light on this issue.

\section{Sample }

The target clouds (listed in Table \ref{sample}) comprise a subset of the 
\citet{2016ApJ...831...73V}
sample, chosen to sample a range of conditions and environments, as well
as for suitable size (8 to 35 pc) and  \added{relative} lack of confusion.
The sample has maps of CO and \coo\ 
\added{from the Five College Radio Astronomy Observatory (FCRAO)
}
and \mm\ continuum emission from
the Bolocam Galactic Plane Survey (BGPS),
\added{obtained at the Caltech Submillimeter Observatory (CSO)
}
\citep{2011ApJS..192....4A, Ginsburg:2013},
and two measures of the
star formation rate, radio continuum and mid-infrared emission
\added{
\citep{2016ApJ...831...73V}.
}
\added{The CO and  \coo\ data were respectively taken from the UMass-Stony Brook Survey \citep{1986ApJS...60....1S}
and the Boston University-FCRAO Galactic Ring
Survey (GRS) 
\citep{2006ApJS..163..145J}.
The FWHM beam sizes were 45\arcsec\ for CO and 46\arcsec\ for \coo.
The velocity resolutions were $0.65$ \kms, smoothed to $1.0$ \kms, for CO 
and $0.21$ \kms\ for \coo. The RMS noise (in \tastar) 
was $0.4$ K for CO and $0.13$ K for
\coo. 
The BGPS was obtained with a filter centered at 271.4 GHz and a bandwidth of
46 GHz (avoiding the CO \jj21\ line) at 33\arcsec\ effective resolution and RMS noise ranging from $11-53$ mJy beam$^{-1}$ \citep{2011ApJS..192....4A}.
We use version 2.0 of the BGPS catalog, which
had improved positional accuracy and response to extended structure
\citep{Ginsburg:2013}. In particular, 95\% of the flux was recovered for
scales between $33\arcsec$ and $80\arcsec$, and emission out to 300\arcsec\ was
partially recovered.
}

The typical angular extent of BGPS sources is a few arcmin, so we selected
clouds with \coo\ extents of 10\arcmin\ to 20\arcmin\ to fully sample the ``diffuse" gas.
The boundaries of the \coo\ emission (column 9 of Table \ref{sample})
 were set in order to separate the cloud
from the background; they were substantial and varied from cloud to cloud.
\added{The procedure was described in \citet{2016ApJ...831...73V}, but
generally the threshold was the minimum needed to distinguish the cloud
from background/foreground \coo\ emission.}
These thresholds are 5 to 12 times the RMS noise, so the clouds
undoubtedly are larger and more massive, but confusion in the inner
Galaxy limits the region that can be isolated.
We favored clouds with at least one BGPS
source with a size of at least 3\arcmin\ so that we can
clearly compare the morphology of the HCN/\hcop\ emission and the
dust emission. We also favored clouds with larger and stronger BGPS
sources.
We identified 6 clouds in the sample of
\citet{2016ApJ...831...73V}
that meet these criteria. The sample spans a good range of cloud mass
(2.4\ee4 \msun\ to 3.3\ee5 \msun), dense gas mass (700 to 1.3\ee4 \msun),
and star formation rate (23 to 275 \msunmyr). While the clouds range
in distance from us (3.5 to 10.4 kpc), all lie between 5.1 and 6.4 kpc
from the Galactic center, in the molecular ring.

The distances used by
\citet{2016ApJ...831...73V}
were mostly taken from 
\citet{2014ApJS..212....1A},
who used a variety of sources of information on the velocity
and methods for kinematic distance ambiguity resolution (KDAR).
We now have velocities of the dense gas traced best by \hcopi\ or
\hcop\ (see later section) and there are newer distance estimators.
We recalculated the distances using the tool described in
\citet{2018ApJ...856...52W},
which provides a distance pdf and two-sided uncertainties.
While we use the two-sided uncertainties, they are in general
fairly symmetric.
We use the same choice of near, far, or tangent point distances
as \citet{2014ApJS..212....1A}
(noted by N, F, or T in the KDAR column of Table \ref{sample}). 
 In particular, 
\cloudd\ was placed at the tangent point because its velocity was
within 10 \kms\ of the velocity at the tangent point. We have
scaled the sizes, cloud masses, dense gas masses, and star formation rates
to the new distances and propagated the uncertainties in the distance to
uncertainties in other quantities. The distance uncertainties typically
dominate if they enter the calculation of a quantity.
We use the star formation rates from the mid-infrared emission.
The results are in  Table \ref{sample}.
As discussed in detail in the Appendix, we use the kinematic distance
to \clouda\ rather than the closer distance from maser parallax.
The latter is quite uncertain and would require an unreasonably large
(40 \kms) peculiar motion.

\begin{table*}[h] 
\centering
\caption{Sample of Clouds} \label{sample} 
\vspace {3mm} 
\begin{tabular}{l c c c c c c c c r } 
\tableline 
\tableline 
Source  & $d$ & KDAR & Size & Log \mcloud\ & Log \mdense\ &  Log SFR & Map Size & \coo\ Lim &  Note \cr 
        &  (kpc) &  & (pc)  & (\msun)    & (\msun) & (\msun/Myr)  & ($\arcmin \times \arcmin$) & (K)  & \cr 
\tableline 
\clouda\  & $3.48^{+0.43}_{-0.32}$ & N & $22.68^{+2.80}_{-2.09}$ & $4.75^{+0.17}_{-0.25}$ & $4.13^{+0.10}_{-0.09}$  & $2.44^{+0.10}_{-0.10}$ & $30\times30$ & 2.2 & \blank \cr 
\cloudb\  & $10.43^{+0.38}_{-0.41}$ & F & $35.41^{+1.29}_{-1.39}$ & $5.52^{+0.15}_{-0.23}$ & $4.08^{+0.15}_{-0.23}$  & $1.85^{+0.05}_{-0.06}$ & $10\times20$ & 2.6 & 1 \cr 
\cloudc\  & $8.68^{+0.56}_{-0.60}$ & F & $17.21^{+1.11}_{-1.19}$ & $4.96^{+0.15}_{-0.24}$ & $3.49^{+0.15}_{-0.23}$  & $1.36^{+0.07}_{-0.08}$ & $14\times22$ & 2.2 & 1 \cr 
\cloudd\  & $6.60^{+0.13}_{-0.14}$ & T & $27.15^{+0.53}_{-0.58}$ & $5.45^{+0.15}_{-0.23}$ & $3.02^{+0.16}_{-0.25}$  & $2.09^{+0.04}_{-0.05}$ & $10\times26$ & 1.9 & \blank \cr 
\cloude\  & $8.31^{+0.46}_{-0.62}$ & F & $25.69^{+1.42}_{-1.92}$ & $5.41^{+0.15}_{-0.24}$ & $3.91^{+0.15}_{-0.24}$  & $1.87^{+0.06}_{-0.09}$ & $30\times30$ & 1.6 & \blank \cr 
\cloudf\  & $3.71^{+0.68}_{-0.60}$ & N & $8.27^{+1.52}_{-1.34}$ & $4.38^{+0.19}_{-0.32}$ & $2.85^{+0.19}_{-0.32}$  & $1.81^{+0.14}_{-0.18}$ & $20\times10$ & 1.6 & \blank \cr 
\tableline 
\end{tabular} 
 
Notes: 1. Mapped in equatorial Coordinates 
\end{table*}

\section{Observations}

We mapped the six clouds in the HCN (1$-$0) (88.631847~GHz) 
\citep{DeLucia69} and HCO$^+$ (1$-$0) (89.188526~GHz) \citep{Sastry81} lines simultaneously using the SEQUOIA array with 16 pixels in a 4$\times$4 array at the 14-m telescope of the Taeduk Radio Astronomy Observatory (TRAO)
\citep{1999PKAS...14..123R, Jeong19}.
The observations were conducted by the On-The-Fly (OTF) technique in the absolute position switching mode with OFF positions checked to be free from appreciable emission. The mapped areas were different for the individual clouds (Table~1). G034.997+00.330 and G036.459$-$00.183 were mapped in  equatorial coordinates occasionally between February and May in 2017, while the others were mapped in  Galactic coordinates between January and April in 2018.  The  backend was a 2G FFT spectrometer that can accept 32 signal streams at the same time. Thus it is possible to observe two transitions simultaneously between 85 and 100~GHz or 100 and 115~GHz. The backend bandwidth is 62.5~MHz with 4096 channels, yielding a velocity resolution of 0.05~km~s$^{-1}$. The observed line temperature was calibrated on the \tastar\ scale by the standard chopper wheel method. We checked the telescope pointing and focus every 3 hours by observing strong SiO maser sources at 86~GHz. The pointing accuracy was better than 10$''$. The system temperatures depended on the weather condition and source elevation. They were usually around 200~K during the observing runs. The RMS noise levels of the observed spectra were typically about 0.1~K after smoothing to 0.2~km~s$^{-1}$ resolution.
We smoothed the data with the boxcar function to a velocity resolution of
about 0.2 \kms\ when fitting line profiles, but some analysis of the data 
cubes was done with the original resolution.
To assess optical depth effects, we observed the innermost footprint
in the lines of H$^{13}$CN and H$^{13}$CO$^+$
\added{with the same method and RMS noise.} 

\added{The TRAO telescope is the same model as the FCRAO telescope. The panels were readusted and the radome was replaced before our observations. The resulting instrumental properties are described in \citet{Jeong19}. The TRAO has a main-beam size of 58\arcsec\ and a main-beam efficiency of 46\% at 89 GHz. The individual SEQUOIA beams vary in beam size (efficiency) by only a few arcseconds (few percent). The beams are very circular; variation of efficiency with elevation is less than 3\%. 

Many of our sources are quite extended, so the efficiency on the Moon is more relevant for some. That has not yet been measured for the TRAO, but it should be at least as high as that of the FCRAO, where $\eta_{\rm Moon} = 0.7$. The error beam has not been measured for the TRAO, but it should be no worse than that of the FCRAO. There are two error beams for the FCRAO: one is caused by the radome and the other by small-scale irregularities on the panels. The radome error beam is spread over 4 degrees at $-30$ dB (0.1\%), too weak and diffuse to be an issue here. The panel error beam has a size of 30\arcmin\ at $-18$ dB ($1.6$\%). We will assess whether this error beam could affect our results in \S \ref{radprofs}.}


\begin{figure*}[ht!]
\centering
\includegraphics[width=1.0\textwidth]{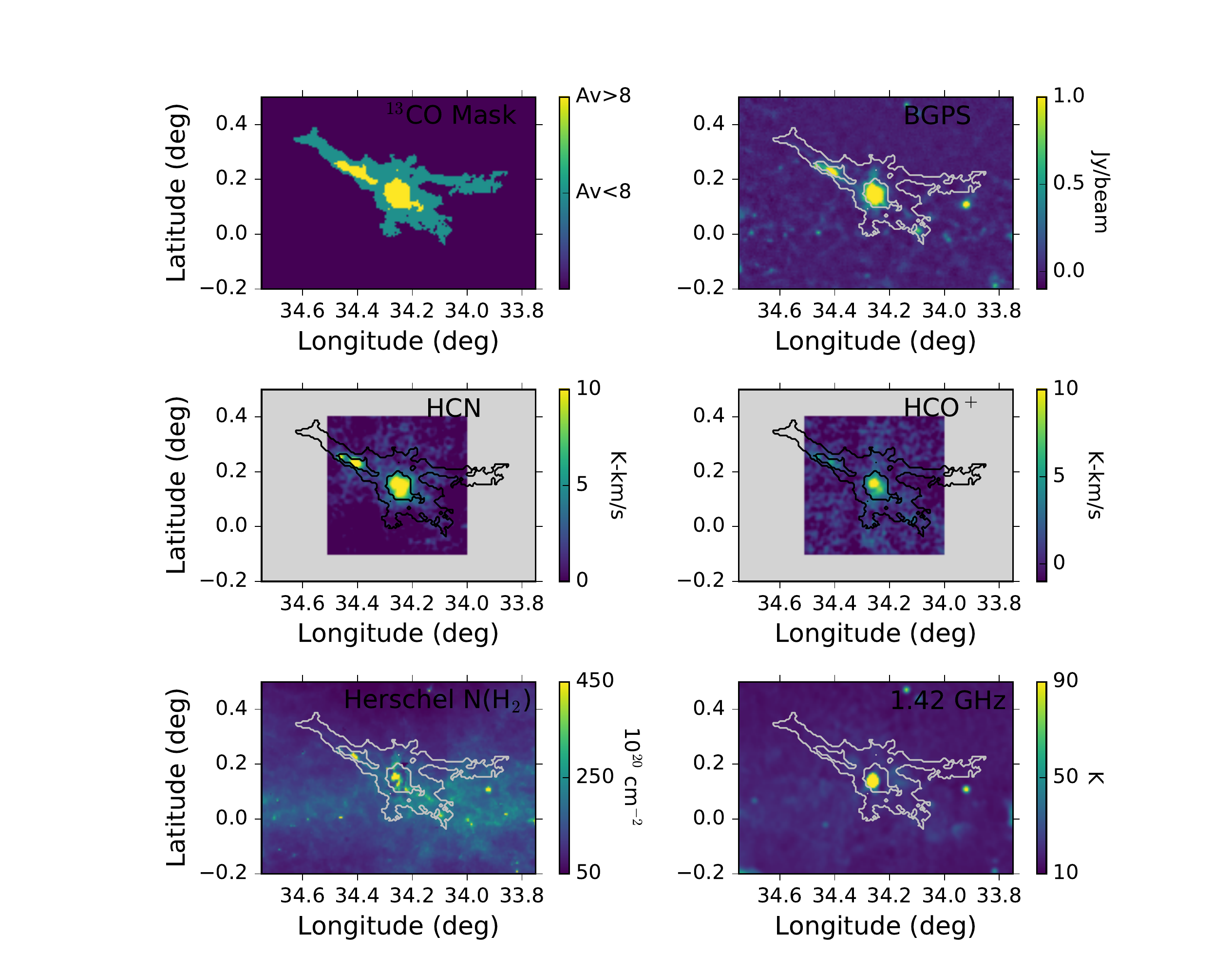}
\caption{This figure is for the cloud \clouda.
Except for the upper left panel, 
the GRS \coo\ contours are shown in black or white, depending
on the panel, along with images of other tracers or masks.
Upper left is the mask for column density $\av \ge 8$ mag and the
outermost contour of \coo\ emission;
upper right is the BGPS \mm\ continuum emission in color with 
the \coo\ contour in white; 
middle left is the HCN integrated intensity in color with 
the \coo\ contour in black;
middle right is the \hcop\ integrated intensity in color with 
the \coo\ contour in black;
lower left is the gas column density determined from Herschel in color
with the \coo\ contour in white;
lower right is the 1.42 GHz radio continuum emission in color
with the \coo\ contour in white;
The size of the box in the middle panels shows the region mapped in HCN and \hcop.
The \coo\ is in units of K \kms ($T_{\rm mb}$) scaled 0 to 50 K \kms;
Color bars indicate the range and scaling of the other tracers.
}
\label{G034.158-2by3}
\end{figure*}

\begin{figure*}[ht!]
\centering
\includegraphics[width=1.00\textwidth]{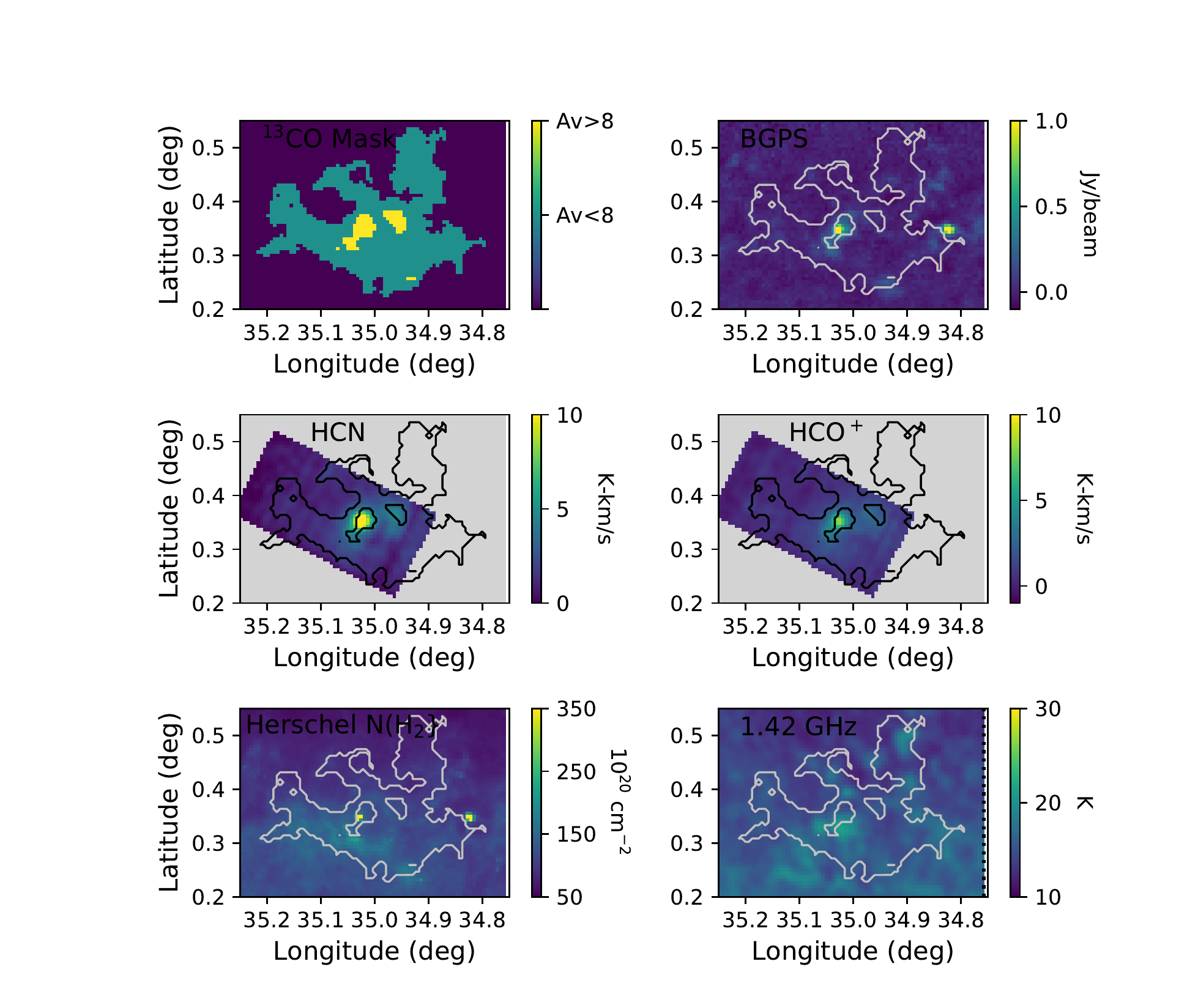}
\caption{This figure is for the cloud \cloudb.
Except for the upper left panel, 
the GRS \coo\ contours are shown in black or white, depending
on the panel, along with images of other tracers or masks.
Upper left is the mask for column density $\av \ge 8$ mag and the
outermost contour of \coo\ emission;
upper right is the BGPS \mm\ continuum emission in color with 
the \coo\ contour in white; 
middle left is the HCN integrated intensity in color with 
the \coo\ contour in black;
middle right is the \hcop\ integrated intensity in color with 
the \coo\ contour in black;
lower left is the gas column density determined from Herschel in color
with the \coo\ contour in white;
lower right is the 1.42 GHz radio continuum emission in color
with the \coo\ contour in white;
The size of the box in the middle panels shows the region mapped in HCN and \hcop.
The \coo\ is in units of K \kms ($T_{\rm mb}$) scaled 0 to 50 K \kms;
Color bars indicate the range and scaling of the other tracers.
}
\label{G034.997-2by3}
\end{figure*}

\begin{figure*}[ht!]
\centering
\includegraphics[width=1.00\textwidth]{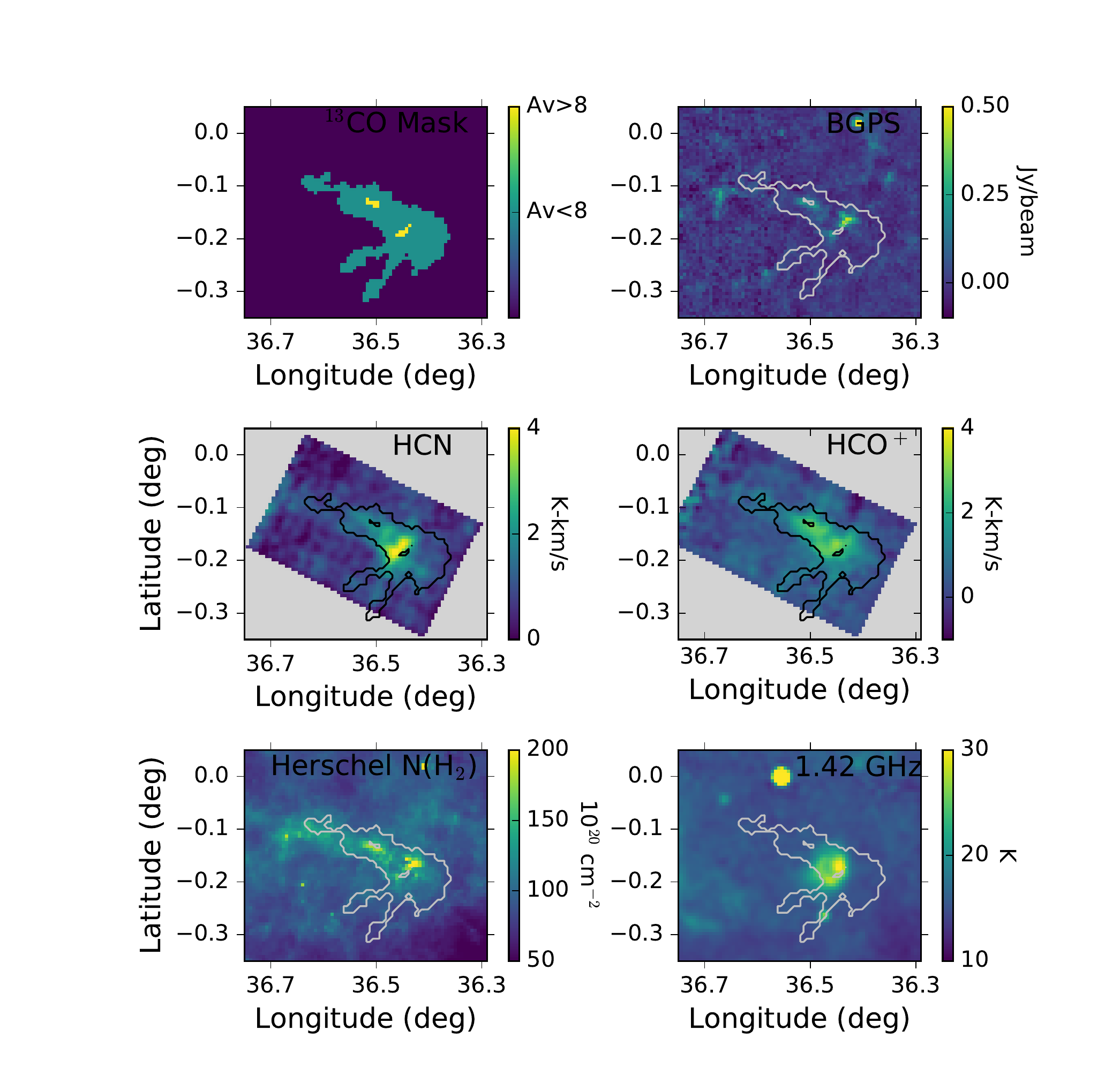}
\caption{This figure is for the cloud \cloudc.
Except for the upper left panel, 
the GRS \coo\ contours are shown in black or white, depending
on the panel, along with images of other tracers or masks.
Upper left is the mask for column density $\av \ge 8$ mag and the
outermost contour of \coo\ emission;
upper right is the BGPS \mm\ continuum emission in color with 
the \coo\ contour in white; 
middle left is the HCN integrated intensity in color with 
the \coo\ contour in black;
middle right is the \hcop\ integrated intensity in color with 
the \coo\ contour in black;
lower left is the gas column density determined from Herschel in color
with the \coo\ contour in white;
lower right is the 1.42 GHz radio continuum emission in color
with the \coo\ contour in white;
The size of the box in the middle panels shows the region mapped in HCN and \hcop.
The \coo\ is in units of K \kms ($T_{\rm mb}$) scaled 0 to 50 K \kms;
Color bars indicate the range and scaling of the other tracers.
}
\label{G036.459-2by3}
\end{figure*}

\begin{figure*}[ht!]
\centering
\includegraphics[width=1.00\textwidth]{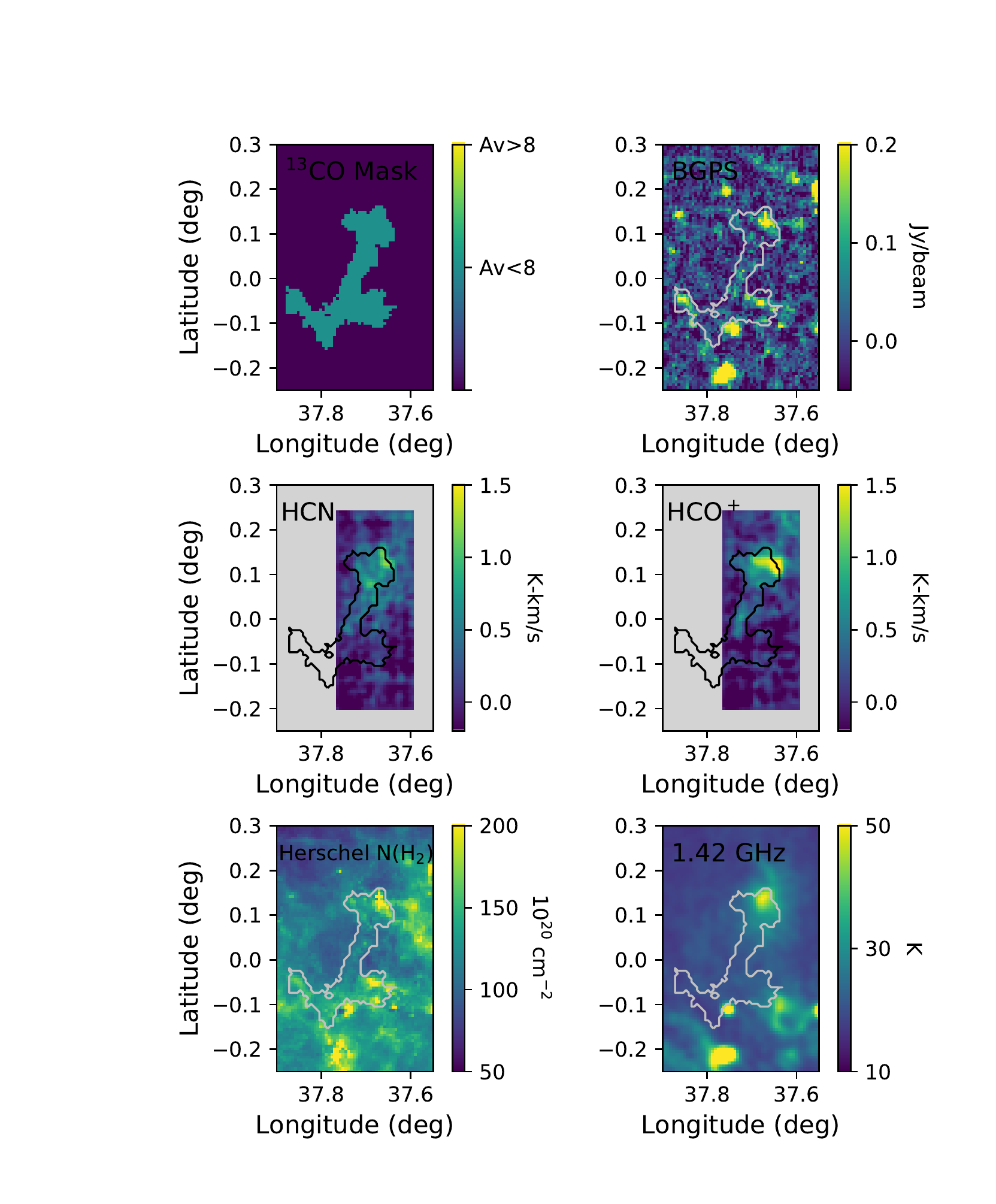}
\caption{This figure is for the cloud \cloudd.
Except for the upper left panel, 
the GRS \coo\ contours are shown in black or white, depending
on the panel, along with images of other tracers or masks.
Upper left is the mask for column density $\av \ge 8$ mag and the
outermost contour of \coo\ emission;
upper right is the BGPS \mm\ continuum emission in color with 
the \coo\ contour in white; 
middle left is the HCN integrated intensity in color with 
the \coo\ contour in black;
middle right is the \hcop\ integrated intensity in color with 
the \coo\ contour in black;
lower left is the gas column density determined from Herschel in color
with the \coo\ contour in white;
lower right is the 1.42 GHz radio continuum emission in color
with the \coo\ contour in white;
The size of the box in the middle panels shows the region mapped in HCN and \hcop.
The \coo\ is in units of K \kms ($T_{\rm mb}$) scaled 0 to 50 K \kms;
Color bars indicate the range and scaling of the other tracers.
}
\label{G037.677-2by3}
\end{figure*}

\begin{figure*}[ht!]
\centering
\includegraphics[width=1.00\textwidth]{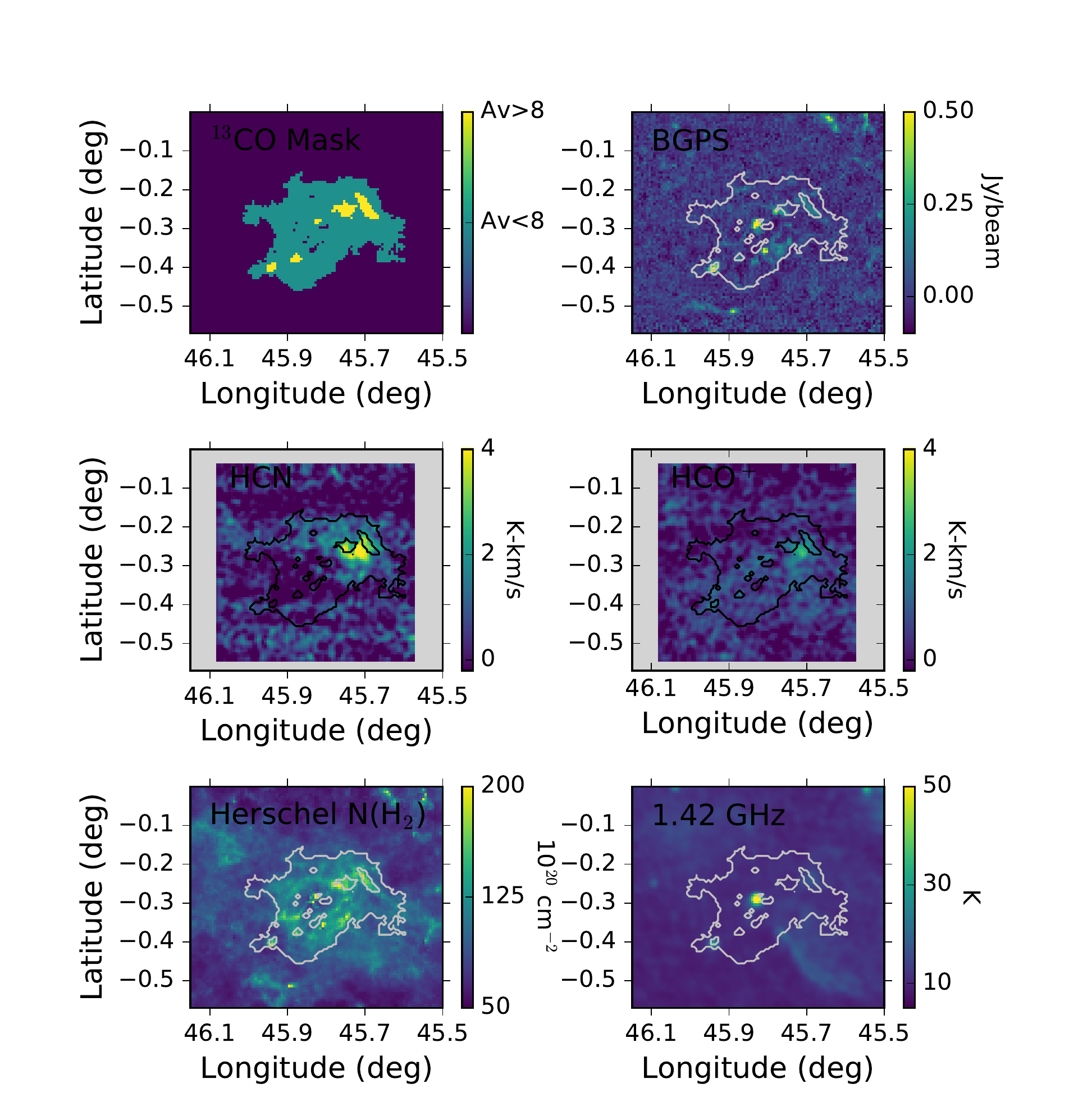}
\caption{This figure is for the cloud \cloude.
Except for the upper left panel, 
the GRS \coo\ contours are shown in black or white, depending
on the panel, along with images of other tracers or masks.
Upper left is the mask for column density $\av \ge 8$ mag and the
outermost contour of \coo\ emission;
upper right is the BGPS \mm\ continuum emission in color with 
the \coo\ contour in white; 
middle left is the HCN integrated intensity in color with 
the \coo\ contour in black;
middle right is the \hcop\ integrated intensity in color with 
the \coo\ contour in black;
lower left is the gas column density determined from Herschel in color
with the \coo\ contour in white;
lower right is the 1.42 GHz radio continuum emission in color
with the \coo\ contour in white;
The size of the box in the middle panels shows the region mapped in HCN and \hcop.
The \coo\ is in units of K \kms ($T_{\rm mb}$) scaled 0 to 50 K \kms;
Color bars indicate the range and scaling of the other tracers.
}
\label{G045.825-2by3}
\end{figure*}

\begin{figure*}[ht!]
\centering
\includegraphics[width=1.00\textwidth]{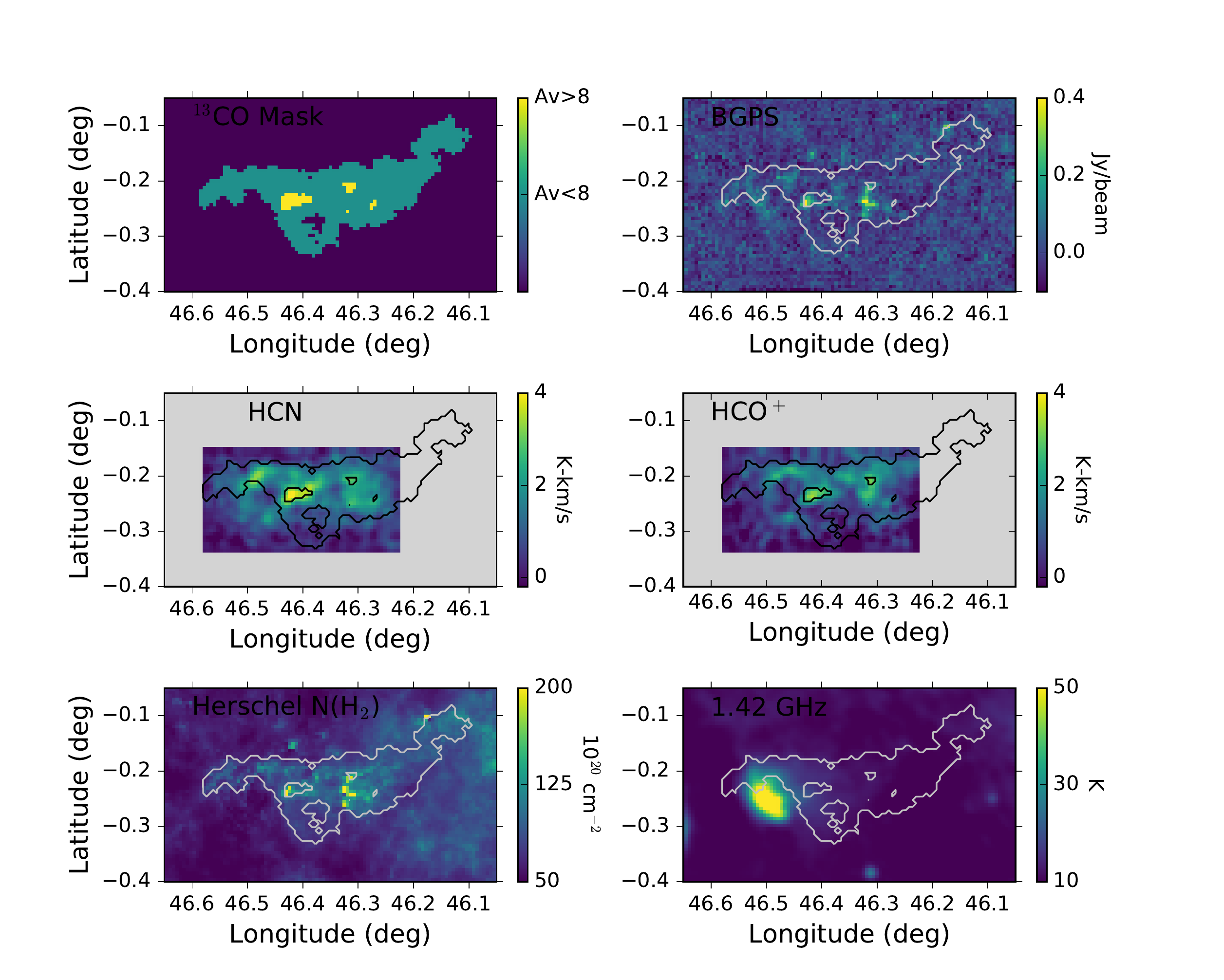}
\caption{This figure is for the cloud \cloudf.
Except for the upper left panel, 
the GRS \coo\ contours are shown in black or white, depending
on the panel, along with images of other tracers or masks.
Upper left is the mask for column density $\av \ge 8$ mag and the
outermost contour of \coo\ emission;
upper right is the BGPS \mm\ continuum emission in color with 
the \coo\ contour in white; 
middle left is the HCN integrated intensity in color with 
the \coo\ contour in black;
middle right is the \hcop\ integrated intensity in color with 
the \coo\ contour in black;
lower left is the gas column density determined from Herschel in color
with the \coo\ contour in white;
lower right is the 1.42 GHz radio continuum emission in color
with the \coo\ contour in white;
The size of the box in the middle panels shows the region mapped in HCN and \hcop.
The \coo\ is in units of K \kms ($T_{\rm mb}$) scaled 0 to 50 K \kms;
Color bars indicate the range and scaling of the other tracers.
}
\label{G046.495-2by3}
\end{figure*}

\section{Results}\label{results}

The most compact presentation of the results is in figures 
\ref{G034.158-2by3} to \ref{G046.495-2by3}. In each figure, 
all panels present the outermost contour of the
velocity integrated intensity of \coo\ emission
 with four tracers of dense gas (in color)
superimposed: 
the $\av > 8$ mag mask based on the \coo\ column density;
the \mm\ continuum emission from BGPS; the HCN integrated intensity; 
and the \hcop\ integrated intensity. (The method used to determine
the $\av >8$ mask is described in \S \ref{tracers}.) Two panels show in color the
column density determined from Herschel data and the radio continuum emission.

Inspection reveals
a considerable range of behavior. \clouda\ and \cloudb\ show strong emission
from both line tracers and BGPS, and all tracers agree qualitatively on the
location of the dense gas. At the other extreme, \cloudd\ has only very weak
emission from either dense line tracer, both concentrated in the upper right corner
of the \coo\ map; \cloudd\ has weak \coo\ emission and no regions with 
$\av > 8$ mag. \cloudc\ and \cloude\
have weak emission from the dense line tracers that is more extended and poorly
correlated with the BGPS and $\av >8$ mag indicators.
\cloudf\ has three, often overlapping, velocity components that complicate
analysis, but in general it shows moderate agreement among the various
tracers of dense gas.
More complete results for each cloud (spectra at peaks, contour diagrams
of integrated intensity, etc.) are provided in the Appendix.

\section{Analysis}\label{analysis}
 
\subsection{Tracer Comparison}\label{tracers}

We quantify the results from the visual comparison in \S \ref{results},
focusing on the fraction of the luminosity of the line tracers that comes
from regions indicated to be dense, based on extinction or \mm\ continuum
emission.  This section is geared toward understanding how the line
tracers will behave when clouds are observed with spatial resolutions
much greater than 1 pc, as in observations of other galaxies. Thus we do
not separate emission from clumps that differ in spatial or velocity 
location. That analysis is done in \S \ref{clumps}.

First we describe the method used to make the maps of column density
and the $\av > 8$ mag mask.
CO (our convention is that the most common isotope is indicated unless 
otherwise specified) and \coo\ data were used to define column density
maps for each target. 
\deleted{For inner Galaxy clouds, the CO and  \coo\ data were respectively taken from the UMass-Stony Brook Survey \citep{1986ApJS...60....1S}
and the Boston University-FCRAO Galactic Ring
Survey (GRS) 
\citep{2006ApJS..163..145J}.
} 
The procedure to convert 
these data into \coo\ column densities for this paper was largely described by \citet{2013MNRAS.431.1296R}. 
In short, we used equations 1-5 of \citet{2013MNRAS.431.1296R}
to calculate the column density of \coo,
assuming low to moderate optical depths of \coo\ emission and optically thick CO  emission.
Monte Carlo simulations are computed to derive the excitation temperature and \coo\ column density and 
corresponding uncertainties generated by the thermal noise of the data. 
We then converted to molecular hydrogen 
by assuming an isotopic $^{12}$C/$^{13}$C ratio of 45 \citep{2005ApJ...634.1126M}
and fractional abundance of 6000 for CO, based on a recent
determination of the CO abundance by
\citet{2017ApJ...838...66L}. 
Our conversion to extinction of $\av = 1.0\ee{-21} N(\hh)$ is also
provided by 
\citet{2017ApJ...838...66L}. 
\added{Fractionation could decrease the ratio of CO to \coo, which would cause
us to overestimate the column density with our assumed isotopic ratio, but we
have no good way to correct for this, and the effect is not large in these relatively warm clouds.
}

The  much higher threshold of 1 g cm$^{-2}$ proposed by
\citet{2003ApJ...585..850M}
to avoid fragmentation and favor the formation of massive stars
is not probed by \coo\ emission, so we used the column density
determined from Herschel data
\citep{2017MNRAS.471.2730M}
 and available at the HIGAL site:
\url{http://www.astro.cardiff.ac.uk/research/ViaLactea/}.
\added{We do not use the Herschel data for regions of more modest
column density because foreground/background emission strongly
contaminates it. It traces only the highest column density regions
accurately in the inner Galaxy.}

These maps of column density were used to define the regions 
of column density corresponding to various thresholds used to define
``dense" gas.
The velocity interval of \coo\ emission was used to limit the range
of velocities of plausible emission from \hcop; the range was extended
for HCN to account for hyperfine splitting. Then the data cubes for
\coo, BGPS, and Herschel column densities 
were convolved, resampled, and aligned with the
TRAO maps. These were used to measure the luminosity inside and outside
the region defined by the
 criterion described above. This allowed us to determine what fraction of
the HCN and \hcop\ emission arose from ``dense" gas, as defined by that criterion.
In this section, the main beam efficiency has been used to correct to the
scale of \tmb. This procedure may overestimate the brightness for
very extended emission.

The luminosity inside and outside regions defining various 
indicators of dense gas are discussed in the following sections.
The integration of luminosity is limited to the
intersection of the \coo-defined cloud and the mapping box from
the TRAO observations for both HCN and \hcop, but not for \coo.
The equations used to compute the luminosity follow:
\begin{equation}
L_{X,in}  = D^2 \int dv \int d\Omega_{in}  T_X(l,b,v), 
\end{equation}
and
\begin{equation}
L_{X,out}  = D^2 \int dv \int d\Omega_{out}  T_X(l,b,v), 
\end{equation}
where $X$ refers to the tracer, $D$ is the distance in pc, and $\Omega_{in}$, 
 and $\Omega_{out}$ are the solid angle of pixels that satisfy (in) or do not satisfy (out) one of the given conditions 
(column density from \coo, column density from Herschel emission, or overlap
with the mask of emission from the BGPS). In practice, a summed spectrum
was constructed from all pixels that satisfied the conditions.
The uncertainties were calculated as follows:
The RMS noise of the summed spectrum ($RMS_{\rm sum}$) is the quadrature sum of RMS noise values of each pixel that satisfies the threshold condition.
The uncertainty in the luminosity is then
\begin{equation}
RMS_{\rm L} = \delta v \sqrt{N_{\rm ch}}  RMS_{\rm sum} d\Omega D^2
\end{equation}
where $\delta v$ is the channel width and 
$N_{\rm ch}$ is the number of channels in the integration range.
In most sources, these uncertainties were quite small compared to
the luminosity, reflecting the fact that they included
only the uncertainties in the summed spectrum. The distance uncertainties
(Table \ref{sample}) were added in quadrature for all luminosities, 
but not for ratios of luminosities, for which the distance cancels out.

\subsubsection{Line Tracers versus the Extinction Criterion}

How much of the luminosity of the line tracers arises within
regions satisfying the extinction criterion ($\av > 8$ mag)? While our
main focus is on HCN and \hcop, we consider \coo\ as well because its
emission is stronger and easier to obtain for other galaxies.

We used the procedure described above
to determine the luminosity of each tracer for regions with column
density above and below that threshold, as measured by the \coo\
column density. The values for the fraction of pixels inside the \av\ criterion,
the log of the total line luminosity, and the fraction of the total arising
inside the $\av > 8$ mag region ($\fl = L_{\rm in}/L_{\rm tot}$) 
are listed for  HCN, \hcop, and \coo\ in Table
\ref{tabav8}. 
 Two-sided errors are given for the logarithmic luminosities, which
are dominated by the distance uncertainties. The distance does not enter
in the \fl\ values, so the uncertainties are symmetric, much smaller, and given in 
parentheses. The mean, standard deviation, and median are given for
the relevant columns.
Since \cloudd\ had no pixels above the criterion, its value for
\fl\ is zero for all three tracers. 
The values of \fl\ vary widely, with \fl\ between 0 and 0.54
for HCN, between 0 and 0.56 for \hcop, and between 0 and 0.33 for \coo. 
HCN and \hcop\ give very similar
results for \fl. In terms of the total line luminosities, 
the mean of the logarithms of \ltot\ is higher by
0.25 for HCN compared to \hcop. So, the two dense line tracers provide similar measures in this sample,
but the luminosity of HCN is somewhat higher than that of \hcop. 
\coo\ provides still higher luminosity, but  worse correlation
with the extinction criterion for the first two clouds and similar
for the others.

\begin{table*}[h]
\centering 
\caption{Line Luminosities versus $\av > 8$} \label{tabav8} 
\vspace {3mm} 
\begin{tabular}{l c c c c c c c} 
\tableline 
\tableline 
Source  & $N/N_{\rm tot}$  & Log $L_{\rm tot}$ & \fl  &         Log $L_{\rm tot}$ & \fl & Log $L_{\rm tot}$ & \fl \cr 
        &  &  HCN   & HCN    & \hcop\ & \hcop\  & \coo & \coo     \cr 
\tableline 
 G034.158+00.147 &  0.163 & $2.68^{+0.10}_{-0.09}$ &  0.539 (0.003) &            $2.54^{+0.10}_{-0.09}$ &  0.559 (0.003) &            $3.75^{+0.10}_{-0.09}$ & 0.332 (0.000) \cr 
 G034.997+00.330 &  0.079 & $3.37^{+0.03}_{-0.04}$ &  0.251 (0.002) &            $3.09^{+0.03}_{-0.04}$ &  0.243 (0.002) &            $4.34^{+0.03}_{-0.04}$ & 0.125 (0.000) \cr 
 G036.459-00.183 &  0.004 & $2.95^{+0.06}_{-0.07}$ &  0.007 (0.000) &            $2.81^{+0.05}_{-0.07}$ &  0.010 (0.000) &            $3.86^{+0.05}_{-0.06}$ & 0.019 (0.000) \cr 
 G037.677+00.155 &  0.000 & $2.22^{+0.02}_{-0.02}$ &  0.000 (0.000) &            $2.02^{+0.02}_{-0.02}$ &  0.000 (0.000) &            $3.59^{+0.02}_{-0.02}$ & 0.000 (0.000) \cr 
 G045.825-00.291 &  0.054 & $3.13^{+0.05}_{-0.07}$ &  0.148 (0.003) &            $2.56^{+0.05}_{-0.07}$ &  0.169 (0.008) &            $3.88^{+0.05}_{-0.07}$ & 0.113 (0.001) \cr 
 G046.495-00.241 &  0.035 & $2.36^{+0.14}_{-0.17}$ &  0.088 (0.002) &            $2.20^{+0.14}_{-0.17}$ &  0.098 (0.002) &            $3.34^{+0.14}_{-0.17}$ & 0.075 (0.000) \cr 
\tableline 
Mean   &  0.056 & 2.79 & 0.172 & 2.54 & 0.180  & 3.79 & 0.111     \cr 
Std Dev.   &  0.055 & 0.41 & 0.185 & 0.36 & 0.190  & 0.31 & 0.109     \cr 
Median   &  0.045 & 2.82 & 0.118 & 2.55 & 0.133  & 3.80 & 0.094     \cr 
\tableline 
\end{tabular} 
 
Notes: 1. Units of luminosities are \kkms pc$^2$. 
\end{table*}

\begin{figure}[ht!]
\centering
\includegraphics[width=0.60\textwidth, angle=00]{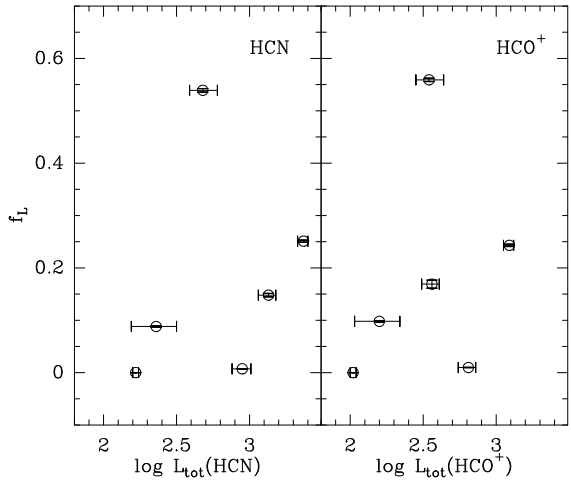}
\caption{The fraction of line luminosity arising above $\av > 8$ mag is plotted
versus total line luminosity. HCN is plotted on the left and \hcop\ on the right.
}
\label{flvsL}
\end{figure}

\begin{figure}[ht!]
\centering
\includegraphics[width=0.60\textwidth, angle=00]{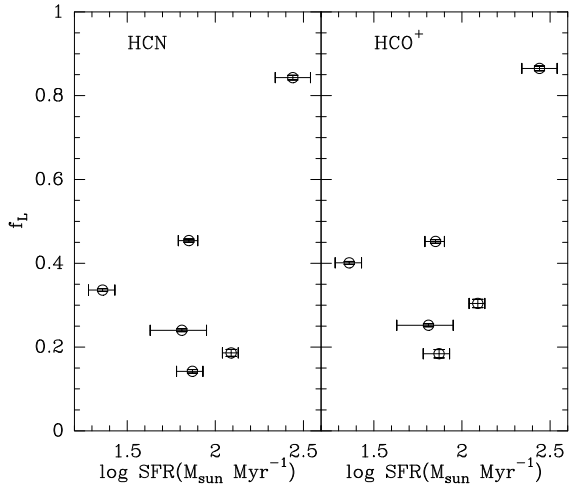}
\caption{The fraction of line luminosity arising in the BGPS emission region is plotted
versus  the star forrmation
rate. HCN is plotted on the left and \hcop\ on the right.
}
\label{flvsSFR}
\end{figure}

Does the fraction of line luminosity arising at high column density correlate
with any observable properties? 
Figure \ref{flvsL} plots \fl\ versus total line luminosity; no trend
is apparent. Neither is there a strong trend of increasing \fl\ with SFR (Fig. \ref{flvsSFR}). While the highest values of \fl\ correspond to the
cloud with the highest star formation rate, the other clouds do not
support an overall trend.

Finally, we comment on the absence of any gas with $\av > 8$ mag in
\cloudd. That cloud is large (27 pc) and massive 
(log $\mcloud = 5.45$), with 
the second highest SFR in the sample. The absence of regions with $\av > 8$ mag
and only weak, fragmented emission in dense gas tracers is thus surprising,
except that the fraction of dense gas indicated by the BGPS emission 
(3.7\ee{-3}) was the lowest in the sample. This cloud is probably more
evolved, with the current episode of star formation coming to an end.

\subsubsection{Line Tracers versus the 1 g \cmc\ Criterion}\label{oneg}

\citet{2003ApJ...585..850M}
proposed a criterion of surface density of 1 g \cmc\ for efficient
formation of massive stars. The \coo\ does not trace such high column densities,
so we used the column density from the Herschel data, as described in \S \ref{tracers}.
There is a small region (10 pixels) in \clouda\ and a single pixel in \cloudb\
that meet the criterion. The fraction of the HCN luminosity in those regions
are 0.034 for \clouda\ and 0.006 in \cloudb. For \hcop, the equivalent
fractions are 0.046 and 0.006; for \coo, they are 0.019 and 0.0.
(For \coo, no pixels were inside that region in \cloudb, presumably because of slightly
different sampling.)
These numbers are interestingly small; the criterion was originally
formulated in the context of the early work on dense clumps identified
with strong localized emission by CS \jj76, 
\citep{1997ApJ...476..730P}
similar in spirit to the HCN and \hcop\ line tracers, but much more biased
towards very high densities.
The sample of 
\citet{1997ApJ...476..730P} was originally based on sources with water masers. As far as we know, \clouda\  is the only source in our sanple with a water maser. 
\added{A recent study of the M17 cloud found that neither HCN nor \hcop\ \jj10\
was particularly good at distinguishing regions above this criterion
\citep{2020ApJ...891...66N}. Higher $J$ transitions are needed to probe such
regions
\citep{2010ApJS..188..313W}.
}

\subsubsection{Line Tracers versus BGPS emission}

\citet{2016ApJ...831...73V}
used the \mm\ continuum emission from the BGPS survey to estimate the
mass of dense gas. To test how well the line tracers correlate with that
criterion for ``dense",
we also computed the line luminosities inside and outside the mask
supplied by the BGPS catalog. In the mask file, each pixel is set to
zero if no emission was detected or to the catalog number of the source if
emission was significant
\citep{Ginsburg:2013}.

\begin{table*}[h]
\centering 
\caption{Line Luminosities versus BGPS Emission} \label{tabbgps} 
\vspace {3mm} 
\begin{tabular}{l c c c c c c c c} 
\tableline 
\tableline 
Source  & $N/N_{\rm tot}$  & Log $L_{\rm tot}$ & \fl  &         Log $L_{\rm tot}$ & \fl & $N/N_{\rm tot}$ & Log $L_{\rm tot}$ & \fl \cr 
        & HCN  &  HCN   & HCN    & \hcop\ & \hcop\ & \coo  & \coo & \coo     \cr 
\tableline 
 G034.158+00.147 &  0.625 & $2.68^{+0.10}_{-0.09}$ &  0.843 (0.006) &            $2.54^{+0.10}_{-0.09}$ &  0.865 (0.006) &            0.612 & $3.75^{+0.10}_{-0.09}$ & 0.741 (0.001) \cr 
 G034.997+00.330 &  0.267 & $3.37^{+0.03}_{-0.04}$ &  0.454 (0.004) &            $3.09^{+0.03}_{-0.04}$ &  0.452 (0.004) &            0.261 & $4.34^{+0.03}_{-0.04}$ & 0.344 (0.000) \cr 
 G036.459-00.183 &  0.205 & $2.95^{+0.05}_{-0.06}$ &  0.336 (0.004) &            $2.81^{+0.05}_{-0.06}$ &  0.401 (0.003) &            0.235 & $3.86^{+0.05}_{-0.06}$ & 0.347 (0.000) \cr 
 G037.677+00.155 &  0.189 & $2.22^{+0.03}_{-0.03}$ &  0.186 (0.008) &            $2.02^{+0.02}_{-0.02}$ &  0.304 (0.010) &            0.230 & $3.59^{+0.02}_{-0.02}$ & 0.275 (0.001) \cr 
 G045.825-00.291 &  0.110 & $3.13^{+0.05}_{-0.07}$ &  0.142 (0.004) &            $2.56^{+0.05}_{-0.08}$ &  0.184 (0.010) &            0.116 & $3.88^{+0.05}_{-0.07}$ & 0.141 (0.001) \cr 
 G046.495-00.241 &  0.145 & $2.36^{+0.14}_{-0.17}$ &  0.240 (0.004) &            $2.20^{+0.14}_{-0.17}$ &  0.252 (0.004) &            0.128 & $3.34^{+0.14}_{-0.17}$ & 0.197 (0.000) \cr 
\tableline 
Mean   &  0.257 & 2.79 & 0.367 & 2.54 & 0.410 & 0.264  & 3.79 & 0.341     \cr 
Std Dev.   &  0.172 & 0.41 & 0.236 & 0.36 & 0.222 & 0.165 & 0.31 & 0.194     \cr 
Median   &  0.197 & 2.82 & 0.288 & 2.55 & 0.353 & 0.233 & 3.80 & 0.309     \cr 
\tableline 
\end{tabular} 
 
Notes: 1. Units of luminosities are \kkms pc$^2$. 
\end{table*}

As can be seen from figure \ref{G045.825-2by3}, \cloude\ 
has very little overlap between the BGPS emission and the HCN/\hcop\ emission.
Consequently, the values in the table for that source are effectively upper limits. For the remaining
four sources, the correlation with BGPS is reasonably strong, as captured in 
\fl\ for BGPS in Table \ref{tabbgps}. 
 As for the extinction criterion, the two dense line
tracers agree well, but the luminosity of HCN is greater
by 0.25 in the log on average.
The \fl\ values for \coo\ are more comparable
to the dense line tracers than was the case for the extinction criterion.
This reflects the fact that the BGPS emission from distant clouds
traces a lower average density than it traces in nearby clouds, so
that it \replaced{corresponds more to}{begins to trace structures between the
scale of} clouds \replaced{than to}{and} dense clumps beyond 
distances of 5-10 kpc
\citep{2011ApJ...741..110D}.
Thus, \coo, HCN, and \hcop\ all seem to trace the gas inferred from 
the BGPS data similarly.
Since
\citet{2016ApJ...831...73V}
used BGPS to measure dense gas mass, we would expect that
their values of SFR per unit
mass would be lower than those for the nearby clouds, where the
$\av > 8$ mag criterion can be used. Indeed, this was the case.

\subsubsection{Conversion of Line Tracer Luminosity to Mass of Dense Gas}\label{conversion}

\begin{figure}[ht!]
\centering
\includegraphics[width=0.60\textwidth, angle=00]{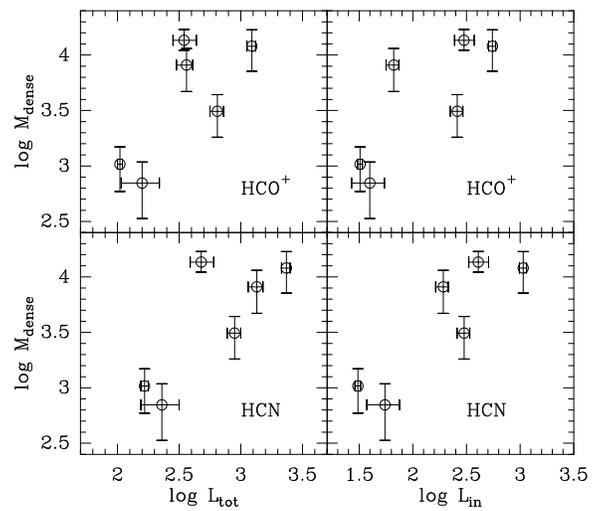}
\caption{The mass of dense gas determined from BGPS data is plotted
versus  the line luminosity. HCN is plotted on the bottom and \hcop\ on the top. The total line luminosities are on the left, the luminosity inside
the BGPS mask on the right.
}
\label{mdvsl}
\end{figure}

We also plot the mass of dense gas determined from BGPS versus the 
luminosity of HCN in figure \ref{mdvsl} for both the total line
tracer luminosity and the line tracer luminosity inside the  BGPS
mask. Both correlate, but the correlation is better for $L_{in}$. 
The values of \mdense, $L_{\rm in}$, and both conversion factors
$\ain = \mdense/L_{\rm in}$ and
$\at = \mdense/L_{\rm tot}$  
 are tabulated in Table \ref{tabmdvsl} for both HCN and \hcop,
along with the means, standard deviations, and medians.

\begin{table*}[h] 
\centering
\caption{\mdense versus Luminosities} \label{tabmdvsl} 
\vspace {3mm} 
\begin{tabular}{l c c c c c c c } 
\tableline 
\tableline 
Source  & Log \mdense & Log $L_{\rm in}$ &  Log \ain &  Log \at & Log $L_{\rm in}$   & Log \ain & Log \at \cr 
        & \msun  &  HCN   & HCN & HCN  & \hcop & \hcop & \hcop     \cr 
\tableline 
 G034.158+00.147 & $4.13^{+0.10}_{-0.09}$ & $2.61^{+0.10}_{-0.09}$ &  $1.52^{+0.02}_{-0.02}$ &  $1.45^{+0.02}_{-0.02}$ & $2.48^{+0.10}_{-0.09}$ & $1.66^{+0.02}_{-0.02}$ & $1.59^{+0.02}_{-0.02}$  \cr 
 G034.997+00.330 & $4.08^{+0.15}_{-0.23}$ & $3.03^{+0.03}_{-0.04}$ &  $1.05^{+0.15}_{-0.22}$ &  $0.71^{+0.15}_{-0.22}$ & $2.74^{+0.03}_{-0.04}$ & $1.34^{+0.15}_{-0.22}$ & $0.99^{+0.15}_{-0.22}$  \cr 
 G036.459-00.183 & $3.49^{+0.15}_{-0.23}$ & $2.48^{+0.05}_{-0.06}$ &  $1.01^{+0.14}_{-0.22}$ &  $0.54^{+0.14}_{-0.22}$ & $2.41^{+0.05}_{-0.06}$ & $1.08^{+0.14}_{-0.22}$ & $0.68^{+0.14}_{-0.22}$  \cr 
 G037.677+00.155 & $3.02^{+0.16}_{-0.25}$ & $1.49^{+0.02}_{-0.03}$ &  $1.53^{+0.16}_{-0.25}$ &  $0.80^{+0.16}_{-0.25}$ & $1.51^{+0.02}_{-0.02}$ & $1.51^{+0.16}_{-0.25}$ & $0.99^{+0.16}_{-0.25}$  \cr 
 G045.825-00.291 & $3.91^{+0.15}_{-0.24}$ & $2.28^{+0.05}_{-0.07}$ &  $1.63^{+0.14}_{-0.22}$ &  $0.78^{+0.14}_{-0.22}$ & $1.82^{+0.05}_{-0.07}$ & $2.09^{+0.14}_{-0.22}$ & $1.35^{+0.15}_{-0.22}$  \cr 
 G046.495-00.241 & $2.85^{+0.19}_{-0.32}$ & $1.74^{+0.14}_{-0.17}$ &  $1.11^{+0.15}_{-0.23}$ &  $0.49^{+0.15}_{-0.23}$ & $1.60^{+0.14}_{-0.17}$ & $1.25^{+0.15}_{-0.23}$ & $0.65^{+0.15}_{-0.23}$  \cr 
\tableline 
Mean   &  3.579 & 2.271 & 1.308 & 0.793 & 2.094 & 1.486 & 1.044  \cr 
Std. Dev.   &  0.505 & 0.521 & 0.254 & 0.315 & 0.471 & 0.326 & 0.339       \cr 
Median   &  3.701 & 2.381 & 1.314 & 0.743 & 2.118 & 1.423 & 0.992       \cr 
\tableline 
\end{tabular} 
 
Notes: 1. Units of luminosities are \kkms pc$^2$. 
\end{table*}

\added{While the BGPS emission is tracing somewhat lower density material
than the $\av > 8$ mag criterion for the distant clouds, it has still been used
to trace dense gas, so a conversion factor between the dense line tracers and
the mass determined from BGPS is of interest.
}
If we restrict the luminosity of HCN to the region 
inside the BGPS mask, we get
$\mean{\log(\ain)} = 1.308\pm .254$, translating to $\mdense
= 20^{+16}_{-9}$ \lhcn. 
Including the \added{luminosity of the} entire cloud, the result is 
$\mean{\log(\at)} = 0.793\pm .315$, translating to $\mdense
= 6.2^{+6.6}_{-3.0}$ \lhcn.
The latter value is more appropriate for extragalactic observations, where
the luminosity will arise from the whole cloud if one wants to estimate
the mass of material with densities similar to that of BGPS sources
in the Galaxy. A conversion factor of 20 (appropriate for the luminosity inside the BGPS
mask) is the same as the average of $20 \pm 5$ derived by 
\citet{2010ApJS..188..313W}
while the value of 6.2 is more similar to, but a bit lower than,
that used in extragalactic studies 
\citep{Gao:2004, 2015ApJ...805...31L, 2019ApJ...880..127J}

\subsubsection{\hcop\ versus HCN versus \coo}

\begin{figure}[ht!]
\centering
\includegraphics[width=0.70\textwidth, angle=00]{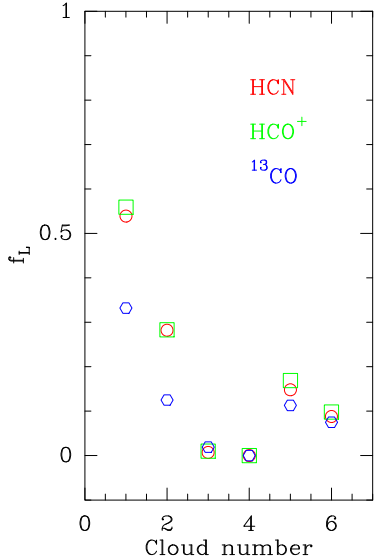}
\caption{The fraction of luminosity (\fl) within the $\av > 8$ mag region
is plotted for each tracer versus cloud number, in the order they
appear in all the tables. Different symbols and colors are used to
separate the different tracers.
}
\label{flvscloud}
\end{figure}

The fraction of luminosity coming from the $\av > 8$ mag region is plotted
for each of the line tracers in figure \ref{flvscloud}.
The first two clouds (\clouda\ and \cloudb) clearly differ from the
other four. The values for \fl\ are higher and those for HCN or \hcop\
clearly exceed those for \coo. For the other 4, all the values are low
and the different tracers are in rough agreement. In those clouds,
the \coo\ would be equally good or bad at tracing the dense gas.

The ratio of total luminosities, $\lhcop/\lhcn$, averaged
in the logs, is 0.56, similar to the 0.7 ratio found in the EMPIRE study,
\citep{2019ApJ...880..127J}
and in the CMZ of the Galaxy
\citep{2012MNRAS.419.2961J}
\added{as well as the ratio of 0.73 found in M17 \citep{2020ApJ...891...66N}.}
but there are substantial variations from cloud to cloud.
Despite the differences in critical density, and probably in chemistry,
the two lines seem to
be tracing similar material on average in this sample.
Both lines trace about the same concentration of emission, as measured
by \fl. For the $\av > 8$ mag criterion, 
the average ratio, $\mean{\fl(\rm \hcop)/\fl(\rm HCN)} = 1.14\pm 0.18$.
This is not what simplistic predictions based on critical densities would
have predicted (see \S \ref{radprofs} and \S \ref{discussion}).

\subsection{Clump Analysis}\label{clumps}

In this section, we search for
peaks, separating different velocity components where necessary. This
analysis is used to characterize the spectra and sizes of the peaks
identified in the line tracers. This analysis is what can be done
with spatial resolution much less than 1 pc, as is common within the
Galaxy.

Our standard procedure is as follows. We examine the data to find all
velocity intervals with significant emission. Then we exclude those regions
while removing a second-order baseline, also using only enough velocity range
to get good baseline on each end ($v_{sp}$) and excluding velocities with
emission ($v_{\rm win}$). 
The values of the total velocity range and excluded
windows are shown in Table \ref{lineprops}. 
Then $I$, \tastar\ integrated over a
velocity range ($v_{\rm I}$) is computed for every mapped position, 
regardless of 
whether a line is detected there. These are plotted in contour diagrams,
which are used to define the peak position, find regions that should be
eliminated, and find the FWHM size. Contour diagrams are made in steps of $2 \sigma$, starting at $2\sigma$,
where $\sigma$ is the RMS noise in the integrated intensity ($I$), 
as listed in 
Table \ref{lineprops}. Spectra at the peak position, or in cases of
weak emission, spectra averaged over nearby positions, were used to 
determine line properties, such as integrated intensity, velocity, 
line width (Table \ref{tablinefits}). This was often an iterative process
in which the line properties were used to refine the velocity range
($v_{\rm I}$) for integrated intensities and the area outside the
2 $\sigma$ emission regions was refined to determine better noise
levels.
\added{The values in Table \ref{lineprops} represent the final values.}

\begin{table*}[h] 
\centering
\caption{Line Fits} \label{tablinefits} 
\vspace {3mm} 
\begin{tabular}{l r r r r r r } 
\tableline 
\tableline 
Source  & Line  &  Intensity & $v$    & $\Delta v$ & RMS(I)  & Note  \cr 
        &       & (\kkms)    & (\kms) & (\kms)     & (\kkms) &       \cr 
\tableline 
G034.158+00.147 & \hcop\ & 10.60 & 56.42 (0.02) &  3.06 (0.05) & 0.48  & \blank \cr 
G034.158+00.147 & HCN    & 9.84  & 56.60 (0.03) &  2.68 (0.05) & 0.48  & \blank \cr 
G034.158+00.147 & \hcopi\ & 3.61 & 57.91 (0.09) &  4.72 (0.22) & 0.28  & \blank \cr 
G034.158+00.147 & \hcni\ & 6.99 & 57.80 (0.11)  &  4.74 (0.25) & 0.53   & \blank \cr 
G034.997+00.330 & \hcop\ & 3.82 & 53.90 (0.03)  &  3.34 (0.09) & 0.23  & \blank \cr 
G034.997+00.330 & HCN    & 7.19 & 54.50 (0.02)  &  3.02 (0.05) & 0.30  & \blank \cr 
G034.997+00.330 & \hcopi\ & 0.42 & 53.27 (0.14) & 2.20 (0.35) & 0.17  & \blank \cr 
G034.997+00.330 & \hcni\ & 1.04 & 53.60 (0.17) &  2.95 (0.39) & 0.34  & \blank \cr 
G036.459-00.183 & \hcop\ & 1.67 & 77.91 (0.06) &  2.56 (0.15) & 0.33  & \blank \cr 
G036.459-00.183 & HCN    & 1.72 & 78.50 (0.11) &  2.94 (0.25) & 0.33  & \blank \cr 
G037.677+00.155 & \hcop & 1.15  & 82.95 (0.07) &  2.52 (0.10) & 0.15   & \blank \cr 
G037.677+00.155 & HCN   & 0.48  & 82.90 (0.21) &  3.93 (0.56) & 0.25   & \blank \cr 
G045.825-00.291 & \hcop\ & 1.31 & 50.45 (0.22) &  5.35 (0.56) & 0.35   & \blank \cr 
G045.825-00.291 & HCN    & 2.23 & 50.18 (0.41) &  11.1 (0.98) & 0.51   & \blank \cr 
G046.495-00.241 & \hcop\ & 1.20 & 51.35 (0.08) &  2.23 (0.20) & 0.11   & v1 \cr
G046.495-00.241 & HCN    & 0.96 & 51.20 (0.14) &  2.75 (0.40) & 0.15   & v1 \cr
G046.495-00.241 & \hcop\ & 0.55 & 54.48 (0.14) &  1.93 (0.34) & 0.10   & v2 \cr
G046.495-00.241 & HCN    & 0.65 & 54.40 (0.16) &  2.27 (0.57) & 0.08   & v2 \cr
G046.495-00.241 & \hcop\ & 1.05 & 58.54 (0.15) &  3.05 (0.45) & 0.16   & v3 \cr
G046.495-00.241 &  HCN   & 1.61 & 58.80 (0.19) &  2.21 (0.36) & 0.16   & v3, hfs \cr
\tableline 
\end{tabular} 
 
Notes: 1. v1 means velocity component 1. 
2. hfs means fit to hyperfine components.
\end{table*}

\subsubsection{What is the origin of the line tracer luminosity from low density regions?}\label{radprofs}

\begin{figure*}[ht!]
\centering
\includegraphics[width=0.90\textwidth]{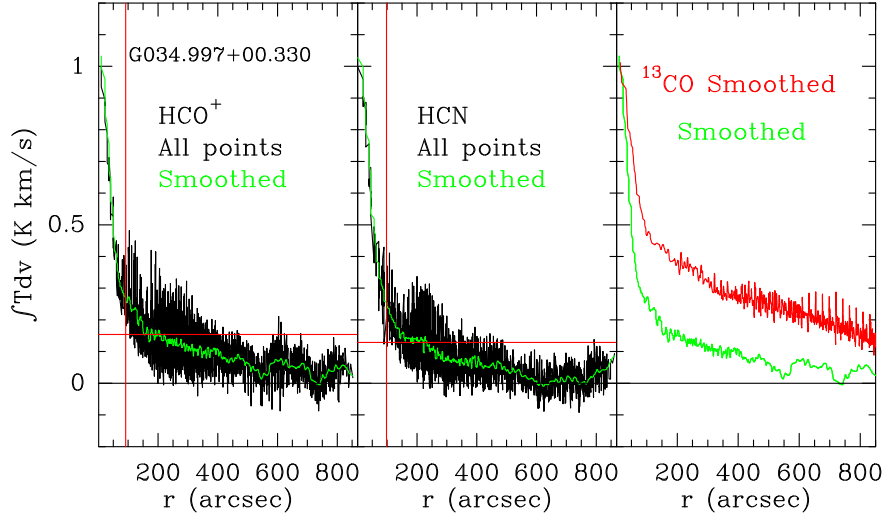}
\caption{The integrated intensity of the \hcop\ and HCN lines
 as a function of distance from the peak position of each for
G034.997+00.330.  
The red vertical line shows the FWHM size that defines the dense core.
The red horizontal line is drawn at 3 times the rms noise in the normalized
intensity.
The panel
on the right shows the \hcop\ and \coo\ lines together, after Hanning 
smoothing by 30 points.
}
\label{g034ivsr}
\end{figure*}

With clear identification of the center of emission, we can explore the origin
of the luminosity of dense line tracers outside the region of the dust tracers.
\cloudb\ provides a good test case because there is a single, clearly defined
peak in the dense line tracers.
In figure \ref{g034ivsr},
the integrated intensities are sorted by distance from the peak
position, normalized to the peak, and plotted versus distance; 
smoothed versions are also plotted.
While both positive and negative values are found at large separations from
the peak, there is a positive
bias. In this source, the roughly 72\% of the dense line tracers' luminosity
arising outside the $\av > 8$ mag mask does not arise in secondary weak emission
peaks, but in very widespread emission that is mostly below the detection
threshold in individual spectra ($\tastar \sim 0.3$ K for a $3 \sigma$
detection).

\added{Could the error beam discussed above create the illusion of low-level, wide-spread emission? Smoothly extended emission at a level of 1.6\% or less relative to the peaked emission could be caused by the error beam. We estimate the observed ratio by using the luminosity inside ($L({\rm in}$)) and outside  ($L({\rm out}$)) the $A_V = 8$ mag contours from Table \ref{tabav8} and the following argument. The surface brightness, $B \propto L/(N \Omega)$, where $N$ is the number of pixels and $\Omega$ is the pixel solid angle. Then
\begin{eqnarray}
R = B({\rm out})/B({\rm in})  & = & L({\rm out})/L({\rm in}) \times N({\rm in})/N({\rm out}) \nonumber \\
 & =  & \fl^{-1} \times N({\rm in})/N({\rm out})
\end{eqnarray}
Approximating $N(\rm{out})$ by $N(\rm{tot})$, since $N(\rm{out}) >> N(\rm{in})$, we use the values in Table 2 to estimate $R$. Values range from 0.29 to 0.40, much larger than the 0.016 that the error beam could contribute. 

The calculation above assumes that the emission is smoothly distributed inside and outside the $A_V = 8$ mag contours. For some sources, the emission is more sharply peaked, so we also divide the average emission in the region outside the $A_V = 8$ mag contours by the peak emission (Table \ref{tablinefits}). These  ratios range from 5\% to 14\%, with the exception of \clouda, for which the ratio is 0.7\%. For all but \clouda, the ratio is {\bf greater} than could be explained by an error beam. The  value for \clouda\ is considerably {\bf less} than the predicted contribution from the error beam, a fact which could be explained if the TRAO error beam is actually lower than that of the FCRAO or by the fact that the \coo\ map did not cover the
full error beam. To summarize, the extended emission from \clouda\ could arise from the error beam, but none of the other clouds' extended emission could so arise.
}

\replaced{Another approach is to}{Having ruled out error beam contributions, we can} use the spectra averaged over the whole region
outside the $\av > 8$ mag region to determine the properties of the
extended emission. For all but \cloudb, the characteristic \tmb\ ranges
from 0.01 K to 0.03 K. 
If this emission is truly distributed, we can 
use RADEX
\citep{2007A&A...468..627V}
 to determine the density of colliders needed to produce such
weak lines. The results are shown in figure \ref{radex}. 
\added{
The dependence of \tmb\ on density is linearly dependent on the density
of colliders in this optically thin, low density regime because every collisional
excitation leads to emission of a photon
\citep{2016ApJ...823..124L}.
}
Because the
observations of \hcop\ and HCN show nearly equal line strengths, we
adjusted the abundance to produce that result. For \hcop, we used
$N({\hcop}) = 1\ee{13}$ \cmc, and for HCN, $N({\rm HCN}) = 2\ee{14}$ \cmc.
For total column density, $N = 1\ee{21}$
\cmc, corresponding to about $\av = 1$ mag, these would correspond to
$X = 1\ee{-8}$ for \hcop, and $X = 2\ee{-7}$ for HCN. Higher total column
densities would allow lower abundances. Such abundances are not 
unreasonable for extinctions of a few, as indicated by figure 8 of
\citet{2017ApJ...841...25G}.
For both, we assumed  $\tk = 20$ K and a linewidth of 1 \kms. These
results illustrate that the low-level, very extended emission that can
dominate the total luminosity can arise in gas with densities as low as
50 to 100 \cmv. These densities are similar to the average densities (\nbar)
of the
entire cloud, determined by dividing the mass by the volume; for the clouds
in this sample, the mean value of this average density,
$\mean{\nbar} = 96 \pm 64$ \cmv
\citep{2016ApJ...831...73V}.
 This is a clear demonstration that common statements
about dense gas tracers are too naive, as discussed further in 
\S \ref{discussion}.

\subsubsection{Properties of the Dense Clumps Identified by the Line Tracers}

\begin{table*}[h]
\centering 
\caption{Clump Properties for HCN} \label{tabclumpshcn} 
\vspace {3mm} 
\begin{tabular}{l c c c c c c r } 
\tableline 
\tableline 
Source  & \rdense\  & \ldense & \fl\  & \mdv\  & $\Sigma$  & $\nbar$  & Note \cr 
        &  (pc)      & (\lunit) &     & (\msun)   & (\msunpc) & (\cmv)  & \cr 
\tableline 
\clouda  & $0.86^{+0.16}_{-0.14}$ & $94.1^{+36.2}_{-32.7}$ & $0.20^{+0.07}_{-0.07}$ & $2845^{+596}_{-548}$ & $1237^{+259}_{-238}$ & $18538^{+7163}_{-6477}$ & \blank \cr 
 \cloudb  & $2.01^{+0.37}_{-0.37}$ & $438.9^{+169.3}_{-169.8}$ & $0.21^{+0.08}_{-0.08}$ & $1454^{+533}_{-533}$ & $114^{+42}_{-42}$ & $728^{+352}_{-352}$ & \blank \cr 
 \cloudc  & $3.61^{+0.39}_{-0.40}$ & $246.4^{+71.2}_{-72.2}$ & $0.28^{+0.08}_{-0.08}$ & $3535^{+563}_{-569}$ & $86^{+14}_{-14}$ & $306^{+75}_{-76}$ & \blank \cr 
 \cloudd  & $1.74^{+0.24}_{-0.24}$ & $18.4^{+7.8}_{-7.8}$ & $0.11^{+0.04}_{-0.04}$ & $1652^{+264}_{-264}$ & $173^{+28}_{-28}$ & $1274^{+367}_{-368}$ & \blank \cr 
 \cloude  & $2.49^{+0.32}_{-0.34}$ & $166.0^{+57.2}_{-59.6}$ & $0.12^{+0.04}_{-0.04}$ & $10631^{+2605}_{-2658}$ & $547^{+134}_{-137}$ & $2816^{+928}_{-970}$ & \blank \cr 
 \cloudf  & $1.12^{+0.25}_{-0.23}$ & $14.4^{+6.7}_{-6.3}$ & $0.06^{+0.03}_{-0.03}$ & $830^{+235}_{-224}$ & $211^{+60}_{-57}$ & $2421^{+1148}_{-1069}$ & v1 \cr 
 \cloudf  & $1.02^{+0.23}_{-0.21}$ & $8.3^{+3.9}_{-3.6}$ & $0.04^{+0.02}_{-0.01}$ & $565^{+235}_{-230}$ & $174^{+73}_{-71}$ & $2199^{+1247}_{-1187}$ & v2 \cr 
 \cloudf  & $1.45^{+0.31}_{-0.28}$ & $37.9^{+16.4}_{-15.1}$ & $0.17^{+0.07}_{-0.06}$ & $2020^{+732}_{-711}$ & $304^{+110}_{-107}$ & $2678^{+1377}_{-1297}$ & v3 \cr 
 \tableline 
Mean   &  1.79 & 128.1  & 0.147 & 2942 &  356 &  3870 & \blank \cr 
Std. Dev.   &  0.86 & 141.7 & 0.075  & 3048 &  360 &  5611 & \blank \cr 
Median   &  1.60 & 66.0 & 0.144 & 1836 &  193 &  2310 & \blank \cr 
\tableline 
\end{tabular} 
 
Note. v1, v2, etc. indicate velocity component when separated. 
\end{table*}

\begin{table*}[h] 
\centering
\caption{Clump Properties for \hcop} \label{tabclumps} 
\vspace {3mm} 
\begin{tabular}{l c c c c c c r } 
\tableline 
\tableline 
Source  & \rdense\  & \ldense & \fl\  & \mdv\  & $\Sigma$  & $\nbar$  & Note \cr 
        &  (pc)      & (\lunit) &     & (\msun)   & (\msunpc) & (\cmv)  & \cr 
\tableline 
\clouda  & $0.55^{+0.14}_{-0.13}$ & $56.2^{+30.3}_{-28.9}$ & $0.16^{+0.09}_{-0.08}$ & $1819^{+487}_{-463}$ & $1936^{+518}_{-492}$ & $45358^{+23133}_{-21888}$ & \blank \cr 
 \cloudb  & $1.82^{+0.36}_{-0.36}$ & $205.7^{+89.0}_{-89.2}$ & $0.17^{+0.07}_{-0.07}$ & $1316^{+494}_{-495}$ & $126^{+47}_{-47}$ & $889^{+454}_{-455}$ & \blank \cr 
 \cloudc  & $6.32^{+0.54}_{-0.56}$ & $681.6^{+177.7}_{-180.9}$ & $1.06^{+0.26}_{-0.28}$ & $6184^{+895}_{-908}$ & $49^{+7}_{-7}$ & $100^{+21}_{-21}$ & \blank \cr 
 \cloudd  & $2.52^{+0.24}_{-0.24}$ & $81.7^{+18.7}_{-18.8}$ & $0.66^{+0.15}_{-0.15}$ & $2389^{+293}_{-294}$ & $120^{+15}_{-15}$ & $609^{+124}_{-124}$ & \blank \cr 
 \cloude  & $3.21^{+0.34}_{-0.37}$ & $150.7^{+51.3}_{-53.5}$ & $0.42^{+0.14}_{-0.14}$ & $13722^{+3213}_{-3286}$ & $424^{+99}_{-101}$ & $1690^{+501}_{-529}$ & \blank \cr 
 \cloudf  & $0.61^{+0.17}_{-0.16}$ & $7.7^{+4.5}_{-4.3}$ & $0.05^{+0.03}_{-0.03}$ & $456^{+151}_{-146}$ & $385^{+127}_{-123}$ & $8045^{+4704}_{-4495}$ & v1 \cr 
 \cloudf  & $1.15^{+0.25}_{-0.23}$ & $8.6^{+4.1}_{-3.8}$ & $0.05^{+0.02}_{-0.02}$ & $638^{+265}_{-259}$ & $154^{+64}_{-62}$ & $1722^{+966}_{-919}$ & v2 \cr 
 \cloudf  & $0.76^{+0.19}_{-0.18}$ & $8.8^{+4.7}_{-4.5}$ & $0.06^{+0.03}_{-0.03}$ & $1058^{+411}_{-400}$ & $580^{+225}_{-220}$ & $9762^{+5701}_{-5446}$ & v3 \cr 
 \tableline 
Mean   &  2.12 & 150.1  & 0.328 & 3448 &  472 &  8522 & \blank \cr 
Std. Dev.   &  1.82 & 211.9 & 0.341  & 4241 &  580 &  14333 & \blank \cr 
Median   &  1.48 & 69.0 & 0.167 & 1568 &  270 &  1706 & \blank \cr 
\tableline 
\end{tabular} 
 
Note. v1, v2, etc.  indicate velocity component when separated. 
\end{table*}

In this section, we derive the properties of the dense clumps, as separated
in position and velocity in order to compare them to the properties
of the dense clumps studied by 
\citet{2010ApJS..188..313W}.
This analysis addresses the question of whether we are comparing apples
to oranges.
For this purpose, we follow the procedure in
\citet{2010ApJS..188..313W}.
This procedure requires the integrated line intensity (I) and the
linewidth ($\Delta v$) at the peak position, the FWHM angular size of the
emission, and the distance.

We use the spectrum of the line at the peak, or in case of weak lines
that are fairly uniformly distributed, an average over spectra surrounding
the peak, to get the intensity and the linewidth.
We use the linewidth from \hcop\ or \hcopi\ also
for the HCN analysis to avoid issues caused by the hyperfine structure
of HCN.
When the lines of the \hcopi\ were strong enough, we used them
to get the linewidth, 
following
\citet{2010ApJS..188..313W}.
This was  possible only for \clouda\ and \cloudb.

We find the angular size of the source at the FWHM of the
integrated intensity map. 
The FWHM source size is determined by plotting the 50\%\ contour and determining
the geometric mean of the two dimensions. This involves some judgment for
weakly peaked regions and some ``peninsulas'' were ignored. 
\added{The uncertainties in the angular size were assessed by repeating the
procedure for the 40\% and 60\% contours; as a result, we included an uncertainty
of 20\arcsec\ for the angular size in the next steps.
}
\deleted{The angular size, along with the distance, determines a linear size
for the half-power of the line tracer emission. 
An uncertainty
of 20\arcsec\ (the pixel size) is used for angular size.
}
\added{The source sizes were generally much larger than the beam size, indicating that beam dilution was a minor effect. The most compact source, \clouda, had a source size of 87\arcsec, much larger than the beam size. While there is undoubtedly structure samller than our beam, it would have a small effect on the following analysis, which is intended to compare our results to those of 
\citet{2010ApJS..188..313W},
who used the FCRAO for \jj10\ observations, hence the same beam size on comparably distant sources.
The method for deriving source sizes was chosen to be similar to that used by 
\citet{2010ApJS..188..313W}.
While crude and somewhat subjective, it provided the most sensible results.
}

This angular size of the source, corrected for beam size,
determines the properties of the dense gas, as defined by the line tracers themselves. 
We use  equation 1 in
\citet{2010ApJS..188..313W} 
to determine the line luminosity (\ldense).
The fraction of line tracer luminosity
inside the region defined by the FWHM of the tracer, extended to a full
Gaussian, was determined by dividing \ldense\ by
the total luminosity in Table \ref{tabbgps}. 

We use equation 3 of
\citet{2010ApJS..188..313W}
to compute  the dense gas mass from the
virial theorem  (\mdv). From \mdv, the 
surface density (\sigmadense), and volume-averaged density (\nbar)
of the gas in the region defined by the line tracer are computed, using
equations 5 and 6 from
\citet{2010ApJS..188..313W}. 
The clump properties determined from HCN are shown in Table
\ref{tabclumpshcn}, while those determined from \hcop\ are shown in Table
\ref{tabclumps}. 
\added{The values in the tables 
have too many significant digits, but the errors given in the tables
 clarify how many
are truly significant.}

The distance uncertainties from Table \ref{sample} are propagated to
other quantities.  The virial mass is determined from the distance and
linewidth; for the linewidth, the value and uncertainty from the line
fits are propagated.
These uncertainties also enter the uncertainties for the surface
density ($\Sigma$) and average density (\nbar) of the clump. Even
with \replaced{likely}{possible} underestimates for the uncertainties in angular size,
the propagated uncertainty in the clump properties is substantial,
especially for \nbar, which depends strongly on the size. Distance
uncertainties were not included in ratios where the distance cancels
out, such as \mdense/\ldense.

The means, standard deviations, and medians are given at the bottom
of each table. The properties vary widely from cloud to cloud, as indicated
by the very substantial standard deviations (larger or comparable to
the mean values). In particular, \cloudc\ has a very large value of
\rdense\ for \hcop, reflecting the very diffuse emission in that tracer; 
the nominal
fraction of luminosity inside the Gaussian is greater than unity,
 while the surface and volume densities are very low.
Clearly, the \hcop\ is not tracing dense gas in this source; HCN
indicates a smaller size and lower \fl, leading to larger surface and
volume densities, but still more characteristic of clouds than of clumps.

For comparison,
\citet{2010ApJS..188..313W}
found a median \rdense\ of
$0.71$, a median \mdense\ of 2.7\ee3 \msun, and
the median \nbar\ of 1.6\ee4 \cmv.
The regions probed by the half-power size of line tracer
emission in the current sample are larger in size, but similar in mass, 
and therefore lower in both surface density and mean density. The
sample of 
\citet{2010ApJS..188..313W}
was derived from studies originally selected by the presence of water
masers, and subsequently, strong emission from CS \jj76, so they probably represented particularly dense regions
\citep{1997ApJ...476..730P}.

Comparing \hcop\ and HCN, the clump properties are broadly similar.
Using the full width of the half power size to define the dense clump
produces similar results for the two tracers in the median, though
differences can be seen in individual sources (e.g., \cloudc). 
While chemical differences
caused by factors like proximity to ionizing sources may well introduce
differences, the two tracers produce similar results in this sample of
Galactic sources.

\subsubsection{Virial Parameters}\label{virial}

Whether or not a particular region is primed to form stars depends
most simply on the relative importance of turbulence versus gravity.
This competition is crudely captured by the virial parameter. However,
measuring the virial parameter is difficult, especially for substructures
within clouds, for which boundaries are somewhat arbitrary and the
contribution of surrounding material is ignored
\citep{2019arXiv191105078M}
It is, however, something observers can estimate.

\begin{table}[ht!] 
\centering
\caption{Virial Parameters} \label{tabvirial} 
\vspace {3mm} 
\begin{tabular}{l c c r } 
\tableline 
\tableline 
Source  & \avd(HCN) & \avd(\hcop)  & Note \cr 
        &           &              &       \cr 
\tableline 
\clouda  & $0.21^{+0.04}_{-0.03}$ & $0.13^{+0.02}_{-0.02}$            & \blank \cr 
 \cloudb  & $0.12^{+0.06}_{-0.06}$ & $0.11^{+0.06}_{-0.06}$            & \blank \cr 
 \cloudc  & $1.13^{+0.47}_{-0.47}$ & $2.01^{+0.83}_{-0.82}$            & \blank \cr 
 \cloudd  & $1.62^{+0.73}_{-0.72}$ & $2.29^{+1.03}_{-1.01}$            & \blank \cr 
 \cloude  & $1.35^{+0.61}_{-0.61}$ & $1.72^{+0.77}_{-0.78}$            & \blank \cr 
 \cloudf  & $4.79^{+2.14}_{-2.11}$ & $3.11^{+1.39}_{-1.37}$            & 1 \cr 
 \tableline 
Mean   &  1.54 &  1.56  & \blank \cr 
Std. Dev.   &  1.56 & 1.10  & \blank \cr 
Median   &  1.24 & 1.86  & \blank \cr 
\tableline 
\end{tabular} 
 
1. Combination of the three velocity components.  
\end{table}

Table \ref{tabvirial} presents the virial parameters, calculated from
$\avd = \mdv/\mdense$. For \cloudf, the values of \mdv\ for the three separate
velocity components have been added together for comparison to \mdense,
which included all the BGPS sources. For some clouds, the BGPS
and dense line tracer maps agree well, but for others,
the agreement is poor (cf. figures \ref{G034.158-2by3} to \ref{G046.495-2by3}). Calculation of \avd\ for those is at best a
crude indicator.

The first two sources have small values for \avd.
The other four sources have $\avd \ge 1$, consistent within uncertainties
with unbound structures and certainly less dominated by gravity.
The spread in \avd\ in this sample is consistent with the range of
values found for BGPS sources in general by
\citet{2016ApJ...822...59S},
with the first two sources lying near the lowest 10\% point
of the distribution (Table 7 of 
\citealt{2016ApJ...822...59S}),
but our median values are higher than those for the full BGPS sample.

\section{Discussion}\label{discussion}

\subsection{The Concept of a Dense Gas Tracer}\label{excitation}

The idea that certain molecules are tracers of dense gas has its origin
in the early days of molecular line astronomy. At that time, sensitivities
were poor, maps were small, and thus the maps of dense gas line tracers
were much smaller than those of CO and \coo. Studies of multiple transitions
of molecules like CS and \form\ provided evidence for gas with densities of
about $n \sim \eten5$ \cmv
\citep{1984ApJ...276..625S,1986ApJ...306..670M,1987ApJ...318..392M}.
These studies and many others led to the naive idea that a single line
of these molecules indicated the presence of gas of a certain density,
often described as the critical density.

The idea of a critical density arose among radio astronomers when most
observations were at centimeter wavelengths, for which the Rayleigh-Jeans
limit is appropriate. In that limit, the excitation temperature of a line
increases from the background temperature of about 2.73 K up the kinetic
temperature over a wide range of densities \added{(2-3 orders of magnitude)}. 
By balancing spontaneous
radiative decay and collisional de-excitation, a ``critical" density can
be defined. As discussed in detail by
\citet{1989RMxAA..18...21E}
the critical density in the R-J limit lies near the low density limit 
\added{of the wide range described above},
just as the excitation temperature begins to rise above the background, 
but the critical density instead lies near the {\it high} density end
of the range, near thermalization, for \mm\ lines, where the R-J
approximation is not valid. Consequently, for \mm\ lines, almost
all emission arises from gas well below the critical density, commonly
called sub-thermal emission. Thus, even for the simplified two-level
molecule, the idea that a particular line arises in gas above the critical
density of that line is incorrect.

Once one drops the two-level approximation, considers collisions to higher
levels, and includes trapping, lines can be appreciably excited at even
lower densities. To make this point,
\citet{1999ARA&A..37..311E}
introduced the concept of the effective density: the density needed to
produce a line of $\tmb = 1$ K. 
\citet{2015PASP..127..299S} explored these issues in greater detail
and computed effective densities (now for the integrated intensity of
1 K \kms) for many transitions, confirming that
many are orders of magnitude less than the corresponding critical densities.
The largest discrepancies between effective and critical densities
occur for the resonance transitions, \jj10, which are the ones used in
extragalactic studies and in this paper.
\added{For example, the effective density for \jj10\ is 9.5\ee2 
\cmv\ versus a critical density of 6.8\ee4 \cmv\ for \hcop; the values
are 8.4\ee3 \cmv\ versus 4.7\ee5 \cmv\ for HCN.
}

\begin{figure}[ht!]
\centering
\includegraphics[width=0.80\textwidth, angle=00]{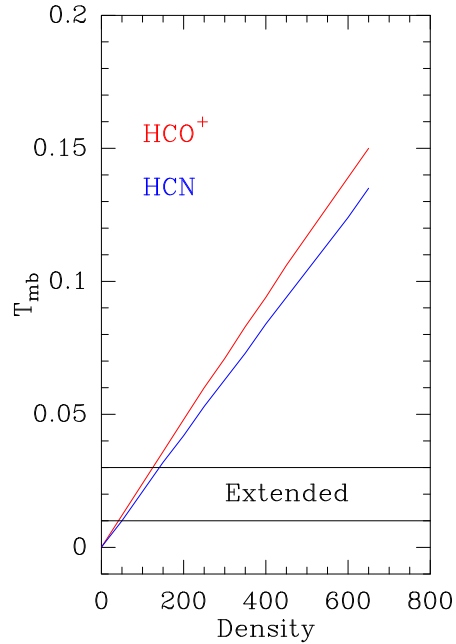}
\caption{The main beam temperature is plotted versus the density
of colliding particles, with \hcop\ in red and HCN in blue.
Typical observations in the regions outside the $\av > 8$ mag
regions are shown as horizontal lines.
}
\label{radex}
\end{figure}

To properly interpret maps of very large regions of clouds in our Galaxy and
observations of other galaxies, it is important to ask
what is the effective density that can produce the distributed emission
\added{that we see, which is at a level much lower than the 1 \kkms\ criterion
used by
\citet{2015PASP..127..299S}.
}
As shown in figure \ref{radex},
for the \jj10\ lines of HCN and \hcop, the entire molecular cloud, at
densities of 50 to 100 \cmv\ and column densities corresponding to 
$\av \approx 1$ can produce \added{the} weak emission \added{that we observe}. 
We have considered only collisions with \hh\ and He. Electron
collisions may be important for low extinction regions of the cloud,
further increasing the emission at low neutral densities
\citep{2017ApJ...841...25G}.
Weak emission ($\tastar = 0.02 -0.1$ K) in
the \hcop\  \jj10\ line from diffuse clouds ($n \approx 100$ \cmv) was observed some
time ago
\citep{1994ApJ...431L.131L}.
In most cases, the area of this
weak emission is large enough that its weak emission dominates the regions
of truly dense gas in determining the total luminosity of the line
tracer. 

We must now ask if there is any remaining validity to using dense gas
tracers. The answer depends on the use we make of them. We cannot say
that they ``probe only the dense gas," but their emission does concentrate into
the dense regions better than CO or \coo\ (see figure \ref{g034ivsr}). 
\added{
Also, lines like \hcop\ \jj10, used in combination with other lines, like
isotopologues of CO, can reveal different regimes of density
\citep{2018A&A...610A..12B}.
}
Star formation rates are
predicted more consistently from nearby clouds to distant galaxies
using the still loosely defined dense gas than from using CO
\citep{2016ApJ...831...73V}. 
Lines from higher $J$ levels will be more strongly biased toward denser gas,
but are harder to observe.

\subsection{Comparison to Other Work}\label{otherwork}

As the \jj10\ transition of 
HCN became more accepted as a probe of dense gas in the extragalactic
community, a number of studies began to examine the origin of HCN (and other
putative tracers of dense gas) in molecular clouds in the Galaxy. We compare
our results to those of other studies in this section.

\citet{2017A&A...599A..98P} mapped a number of molecular lines in Orion B.
They found small fractions of the luminosity of HCN (18\%) and \hcop\ (16\%)
coming from regions with $\av > 15$ mag. They stated that
``The common assumption that lines of large 
critical densities ($\approx \eten5$ \cmv) can only be excited by gas
of similar density is clearly incorrect."
Instead, they conclude that HCN and \hcop\ mostly trace densities from
500 to 1500 \cmv. They note that the tracer that is most strongly
concentrated to the densest gas is \nthp, because of chemistry.

\citet{2017A&A...605L...5K} mapped HCN \jj10\ toward Orion A.
They found that it mainly traced gas with $\av \approx 6$ mag, or
$n \approx 870$ \cmv. They agreed that \nthp\ was the best tracer of truly
dense gas. They also limited the conversion factor: $\mdense \leq 20
\lhcn$, the same as our average value.

\citet{2017A&A...604A..74S} mapped the \jj10\ transitions of 
HCN, \hcop, and their rarer isotoplogues
in the nearby clouds, Aquila, Ophiuchus, and Orion B. They found
that HCN and \hcop\ lines traced the gas down to $\av \approx 2$ or
$n \approx 1\ee3$ \cmv. They found that the conversion factors 
(\at) anti-correlated with the local FUV field strength.

\added{
\citet{2020ApJ...891...66N} studied HCN and \hcop\ \jj10\ toward the M17
cloud. They found that both lines traced equally well regions with 
column density above 3\ee{22} \cmc, but a significant fraction of the
total emission from the cloud came from regions below that column density,
with substantial differences in regions of the cloud with different star
formation histories.
}

\added{
A. Barnes (in prep.) studied the emission from a number of lines toward
the distant ($d = 11.11$ kpc), massive ($\eten4$ to $\eten5$ \msun ) cloud, with
similar results. While HCN \jj10\ emission is strongly enhanced in warm gas
at high column density, a substantial fraction arises in lower column density
gas.
}

Our results are broadly consistent with these works and extend
their conclusions into the inner Galaxy. By considering the far outer
regions of clouds without detections at individual positions, we show that
even lower intensity levels may dominate the total luminosity of 
a cloud in these lines. Simple calculations reveal that even at 
densities of about 100 \cmv,  emission, albeit at very
low intensities, can rival, or even dominate, the more intense emission
from the truly dense regions in detemining the total luminosity of dense line
tracers.

\section{Conclusions}

The main conclusions can be summarized as follows.
\begin{enumerate}
\item The correlation between different tracers of dense gas (extinction,
\mm\ continuum emission, HCN, \hcop) varies from cloud to cloud.
\item Broadly, the clouds divide into two groups, one for which the
dense line tracers are strong and concentrated, and the other for which they
are weak and distributed.
\item In clouds where the dense line tracers are sharply peaked, 
all the tracers show general agreement and are more concentrated than
is the \coo\ emission.
\item In clouds with only weak, distributed emission from dense line 
tracers, the agreement is poor and \coo\ traces similar material.
\item Even when the agreement is good, a substantial fraction of the 
line luminosity arises outside the dust-based measures of dense gas.
\item The agreement of dense line tracers with \mm\ continuum emission
is better than the agreement for $\av > 8$ mag. At the distances of some of
these clouds, the \mm\ continuum emission from BGPS is typically tracing
lower density gas.
\item Measurements of $L(\rm HCN)$ toward other galaxies will likely
include a large fraction of emission from relatively low density gas, 
unless the other galaxy is a starburst galaxy. This variation may be
responsible for some of the observed scatter between galaxies in the
study of
\citet{2019ApJ...880..127J}.
\item The conversion from luminosity to mass of dense gas, as measured
by extinction or \mm\ continuum emission, is quite variable. For the
dense regions, the conversion factor is about 20, while it is closer to
6 if the line luminosity of the whole cloud is included.
\item For this sample, HCN and \hcop\ seem to probe about the same
material. They are equally good (or bad) tracers.
\item The regions probed in this paper in HCN and \hcop\ in two clouds
in the sample are similar to those originally studied by
Wu et al.
but somewhat larger and less dense.
The regions probed in the other clouds are substantially more diffuse
and less clearly bound.
\item The distributed emission of HCN and \hcop\ can arise from regions
of very low density, $n = 50 - 100$ \cmv. Because of the large area of
most clouds at such low densities, these less dense regions can 
dominate the total luminosity of line tracers.

\end{enumerate}

\acknowledgments
We thank the staff of the TRAO for support during the course of these observations. NJE thanks the Department of Astronomy at the University of Texas at Austin for ongoing research support. J.W. is supported by the NSFC grant No. 11590783 and No. 11673029, and National Key R\&D Program of China No.2017YFA0402600.

\appendix

In each section, we discuss the details for each cloud. The
data reduction details are contained in Table \ref{lineprops}.

\begin{table*}[ht!]
\centering
\caption{Reduction Details}
\label{lineprops}
\begin{tabular}{llrrrrrc}
\hline \hline    
Source          & Line   & $v_{\rm sp}$ & $v_{\rm win}$ & $v_{\rm I}$ & Peak Offset & Range         &  Notes \\
                &        & (\kms)       & (\kms)        & (\kms)      & (arcsec)    & (arcsec)      &        \\
\hline
G034.158+00.147 & \hcop\ & 20, 100       & 52, 70        &  52, 70     & 27, 10      &  $-$200, 160; $-$220, 140   &  \\
G034.158+00.147 & HCN    & 20, 100       & 40, 80        &  40, 80     & 20, 4       &  $-$300, 280; $-$280, 240   &  \\
G034.158+00.147 & \hcopi & 20, 100       & 52, 70        &  52, 70     & 31, 29      &  $-$80, 120; $-$80, 140    &  \\
G034.158+00.147 & \hcni  & 20, 100       & 40, 80        &  45, 70     & 27, 20      &  $-$80, 100; $-$80, 100  &  \\
G034.997+00.330 & \hcop\ & 20, 80        & 40, 65        &  40, 65     & $-$20, 110    &  $-$220, 140; $-$160, 220  &  \\
G034.997+00.330 & HCN    & 20, 80        & 42, 63        &  42, 60     & $-$20, 110    &  $-$220, 140; $-$160, 220 &  \\
G034.997+00.330 & \hcopi & 20, 80        & 45, 60        &  45, 60     & $-$20, 110    &   $-$80, 20 ;80, 160    &  \\
G034.997+00.330 & \hcni  & 20, 80        & 42, 63        &  42, 63     & $-$20, 110    &   $-$80, 20 ;80, 160 &  \\
G036.459$-$00.183 & \hcop\ & 20, 120       & 45, 85        &  68, 95     & $-$61, $-$32    &  $-$160, 80; $-$200, 380  &   \\
G036.459$-$00.183 & HCN    & 20, 120       & 43, 93        &  67, 93     & $-$20, $-$20    &  $-$100, 180; $-$160, 340   &  \\
G037.677+00.155 & \hcop\ & 20, 120       & 40, 50; 75, 90 &  77, 88    & $-$98, $-$156   &  $-$180, 120; $-$240, 0   &  \\
G037.677+00.155 & HCN    & 20, 120       & 40, 50; 73, 93 &  73, 93    & $-$40, $-$120   &   $-$160, 240; $-$700, 20 & \\
G045.825$-$00.291 & \hcop\ & 20, 80        & 45, 65        &  45, 65     & $-$418, 60    &  $-$500, $-$180; 20, 180  &  \\
G045.825$-$00.291 & HCN    & 20, 80        & 45, 65        &  40, 65     & $-$400, 80    &  $-$480, $-$220; $-$20, 220  &  \\
G046.495$-$00.241 & \hcop  & 20, 100       & 40,70         &  47, 53     &   87, 3     &  $-$20, 120; $-$60 200    & 1 \\
G046.495$-$00.241 & HCN    & 20, 100       & 40,70         &  47, 53     &  83, 13     &  $-$60, 140; $-$40 80     & 1\\
G046.495$-$00.241 & \hcop\ & 20, 100       & 40,70         &  53, 56     & $-$63, 74     &  $-$140, 340; 0, 120    & 2 \\
G046.495$-$00.241 & HCN    & 20, 100       & 40,70         &  53, 56     & 276, 150    &  $-$140, 340; $-$60,200   & 2 \\
G046.495$-$00.241 & \hcop  & 20, 100       & 40,70         &  56, 60     & $-$307,104    &  $-$540, $-$140; $-$20, 200 & 3 \\
G046.495$-$00.241 & HCN    & 20, 100       & 40,70         &  56, 65     & $-$292,109    &  $-$540, 300; $-$60, 200  & 3 \\
\hline
\end{tabular}

Notes: 
1. Position of center peak, velocity component v1
2. Position of eastern peak, velocity component v2
3. Position of western peak
\end{table*}

\section{\clouda}


\begin{figure}[ht!]
\centering
\includegraphics[width=0.60\textwidth, angle=0]{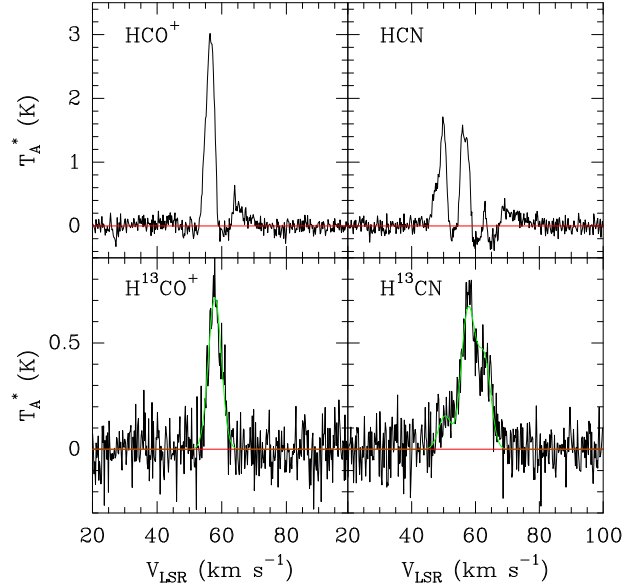}
\caption{The spectra of \hcop, HCN, \hcopi, and \hcni\
toward the peak of \clouda. 
In each plot, the red line shows the zero level. 
For the rare isotopologues, the green line shows the fit to the line.
}
\label{G034.158spec}
\end{figure}

\begin{figure}[ht!]
\centering
\includegraphics[width=0.60\textwidth]{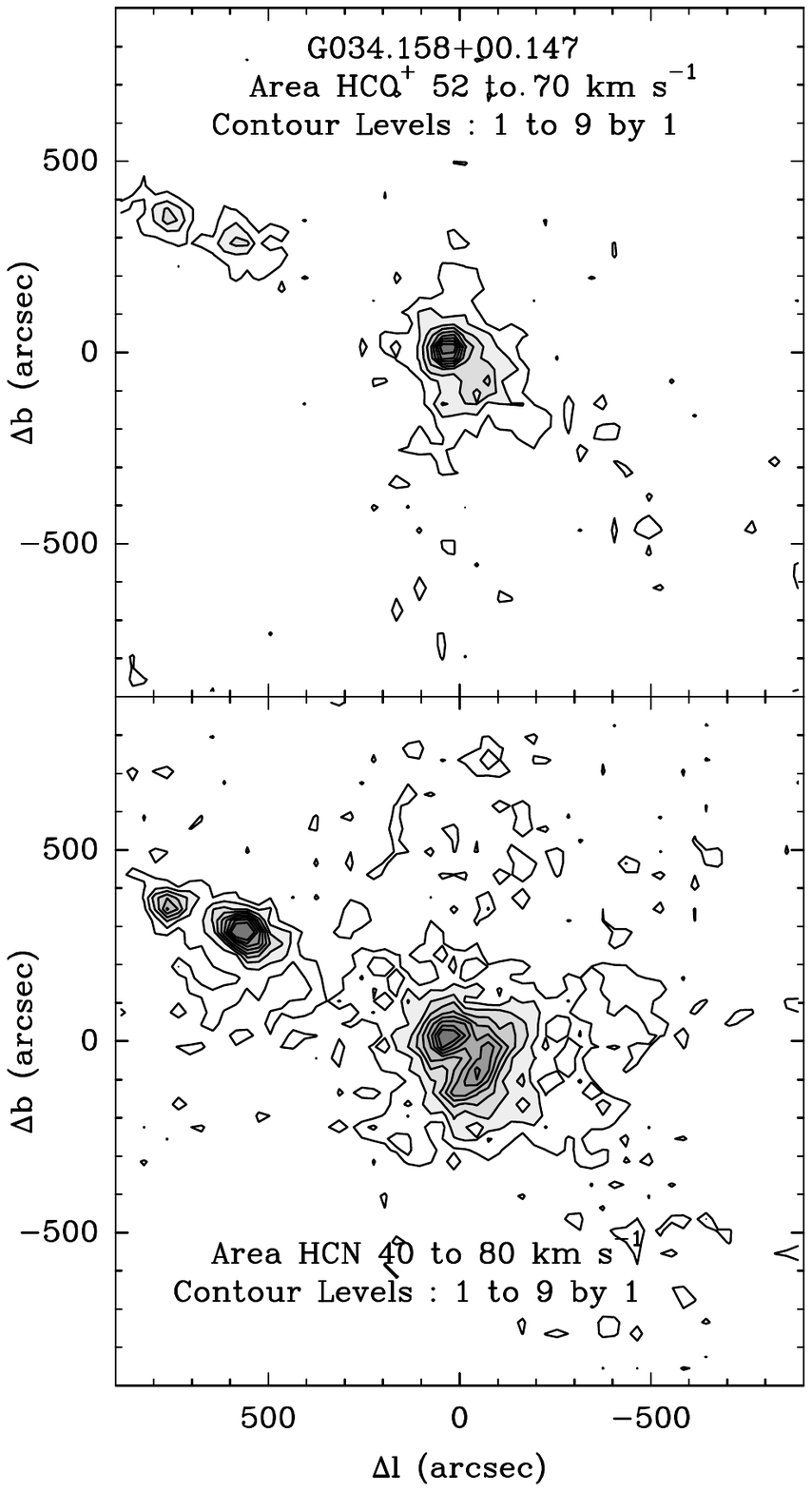}
\caption{The integrated intensities of the \hcop\ (top) and HCN (bottom) lines
  in \clouda. The integration is over the range
of 52 to 70 \kms\ for \hcop\ and 40 to 80 \kms\ for HCN.
}
\label{G034.158maps}
\end{figure}

\begin{figure}[ht!]
\centering
\includegraphics[width=0.60\textwidth]{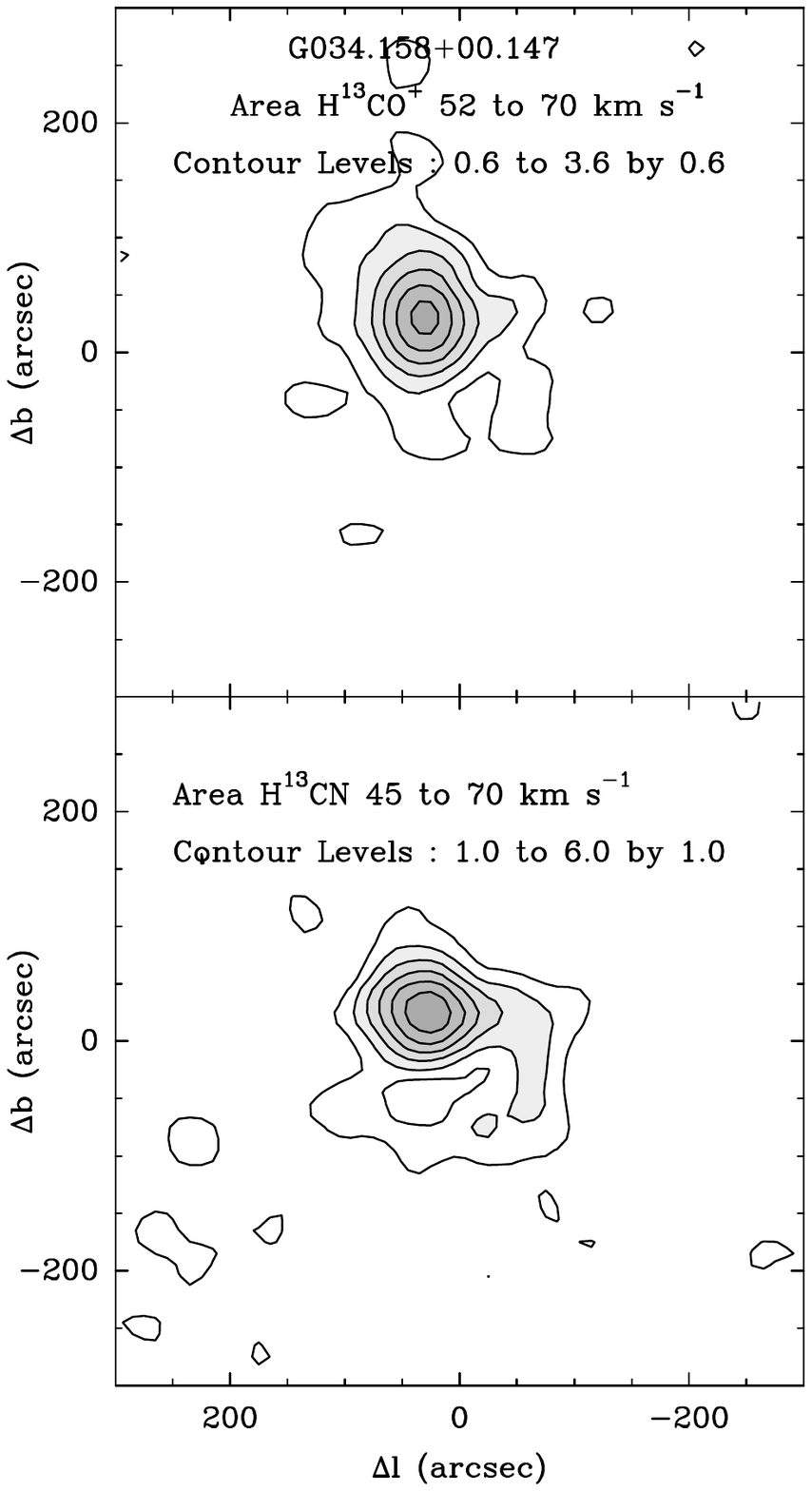}
\caption{The integrated intensities of the \hcopi\ (top) and \hcni\ (bottom) lines
  in \clouda. The integration is over the range
of 52 to 70 \kms\ for \hcopi\ and 40 to 80 \kms\ for \hcni.
}
\label{G034.158-13maps}
\end{figure}

Our map was centered, not on the
source name position, but on $l = 34.250$, $b = 0.150$, near the
\hii\ region, G34.26+0.15. This cloud has been studied extensively under other names.
The clumps to the ``NE” in our maps are associated with the 
IRDC 34.43$+$0.24, in which
\citet{2006ApJ...641..389R}
identified 9 millimeter continuum sources. These are blended into
two clumps in the HCN/\hcop\ maps. G34.26$-$0.15 is a well-known star-forming region
with a water maser \citep{1996A&AS..120..283H} and a methanol maser
\citep{2015MNRAS.450.4109B, 2019ApJS..244....2K}.

The near kinematic distance is 3.7 kpc
\citep{2014ApJS..212....1A}. 
 A VERA parallax measurement
\citep{2011PASJ...63..513K}
of a water maser found by
\citet{2006ApJ...651L.125W}
 gives the distance as $1.56^{+ 0.12}_{-0.11}$ kpc, but
\citet{2011PASJ...63..513K} note that the cloud would
then have a peculiar velocity of about 40 \kms.
Because this seems unlikely, we follow other recent work in
using the kinematic distance.
IRDC G034.43+00.24
has a polarization map 
\citep{2019ApJ...883...95S}.

The BGPS millimeter continuum emission is shown in figure
\ref{G034.158-2by3}.  The maps of integrated intensity peak about
25\arcsec\ east (higher $l$) of the reference position, essentially on
top of the \hii\ region.
Both HCN and \hcop\ show self-absorption and even absorption below
zero around 60 \kms, while the
\hcopi\ spectrum shows a clear peak at somewhat higher velocities
than those of the main lines (Fig. \ref{G034.158spec}),
and the isotopologues peak slightly north of the main lines.
The HCN line is particularly badly affected by absorption.
We attribute this effect to actual absorption of the continuum from
the \hii\ region, which has been subtracted out by the baseline process.
Similar continuum absorption has been seen in \hcop\ \jj10\ and
HCN \jj32\ by 
\citet{2013ApJ...776...29L},
who interpret the line profiles as a signature of infall at about
3 \kms.
\citet{2007ApJ...659..447M}
measured a continuum of 6.7 Jy at 2.8 mm in a source size of 1\farcs6 by
1\farcs4. This would produce a continuum temperature in our 58\arcsec\
beam of $3.5 (2.8/3.4)^a$ K, where $a$ is the spectral index between the two
wavelengths. Because both free-free and dust continuum emission are
contributing in this spectral region, the value of $a$ is uncertain, but
the wavelengths are close enough that it makes little difference. There
is clearly sufficient continuum emission to explain the absorption that we 
observe.
For extragalactic observations, these issues would not be recognized.

To determine core properties, we used the peak integrated intensity from
the fit to the emission and linewidths from fits to the isotopic lines.
For the HCN line properties (Table \ref{tablinefits}), we used the
area within the window of 43 to 70 \kms, because the hyperfine and the
velocity and the absorption made fits impossible. The \hcni\ line for peak 1
was fitted with hyperfine components.

\section{\cloudb}


\begin{figure*}[ht]
\centering
\includegraphics[width=0.60\textwidth,angle=0]{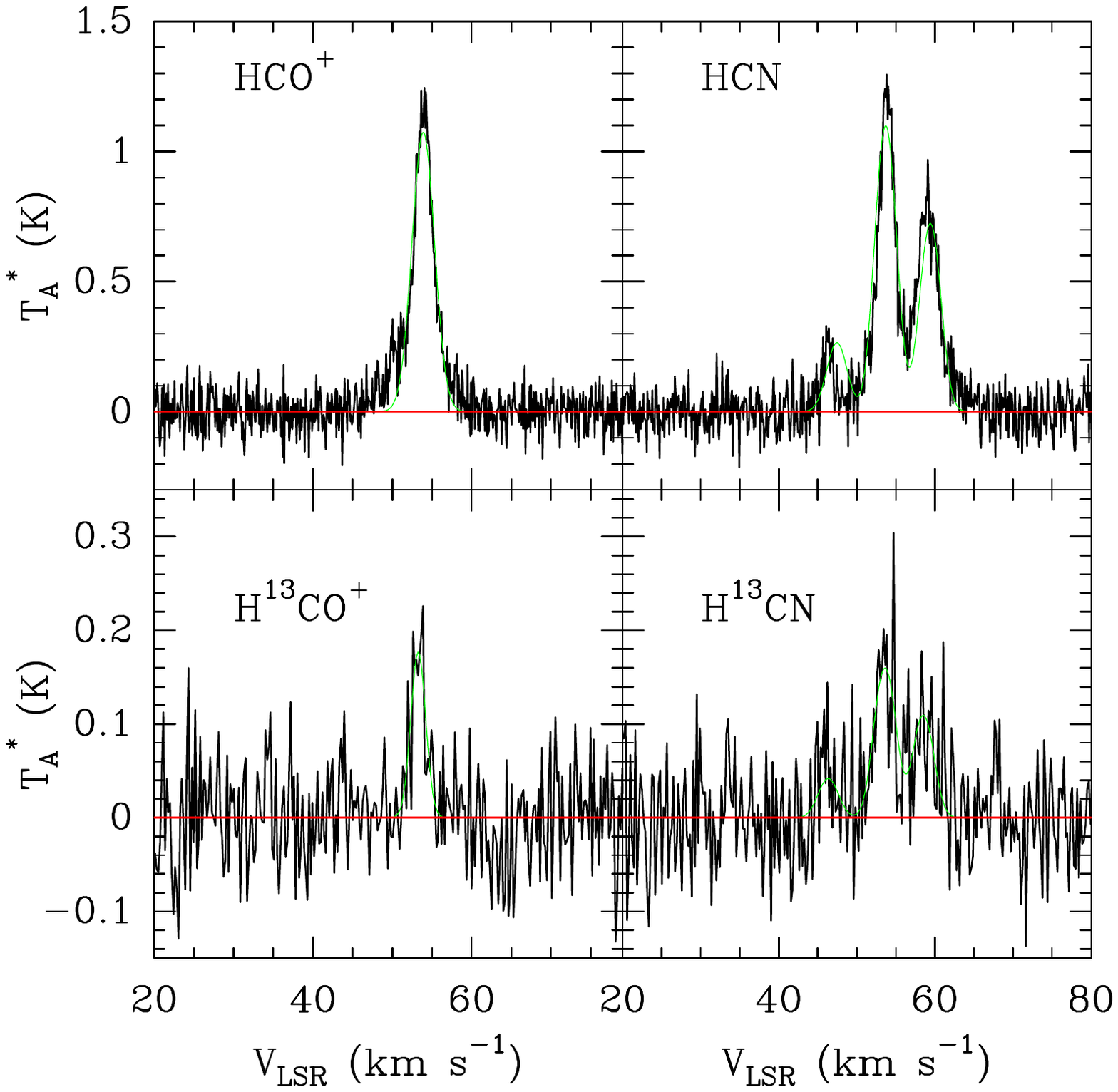}
\caption{Top: The spectra of the \hcop\ and HCN lines at $(-20, 100)$
  for source \cloudb.
 The HCN shows a fit to the hyperfine structure.
 Bottom: The average spectrum of the  \hcopi\ and \hcni\ lines 
 averaged over the nine positions around $(-20, 100)$.
The \hcni\ line was fitted with hyperfine structure.
In each plot, the red line shows the zero level and the green line shows the fit. 
}
\label{G034.997spec}
\end{figure*}

\begin{figure*}[ht]
\centering
\includegraphics[width=0.60\textwidth]{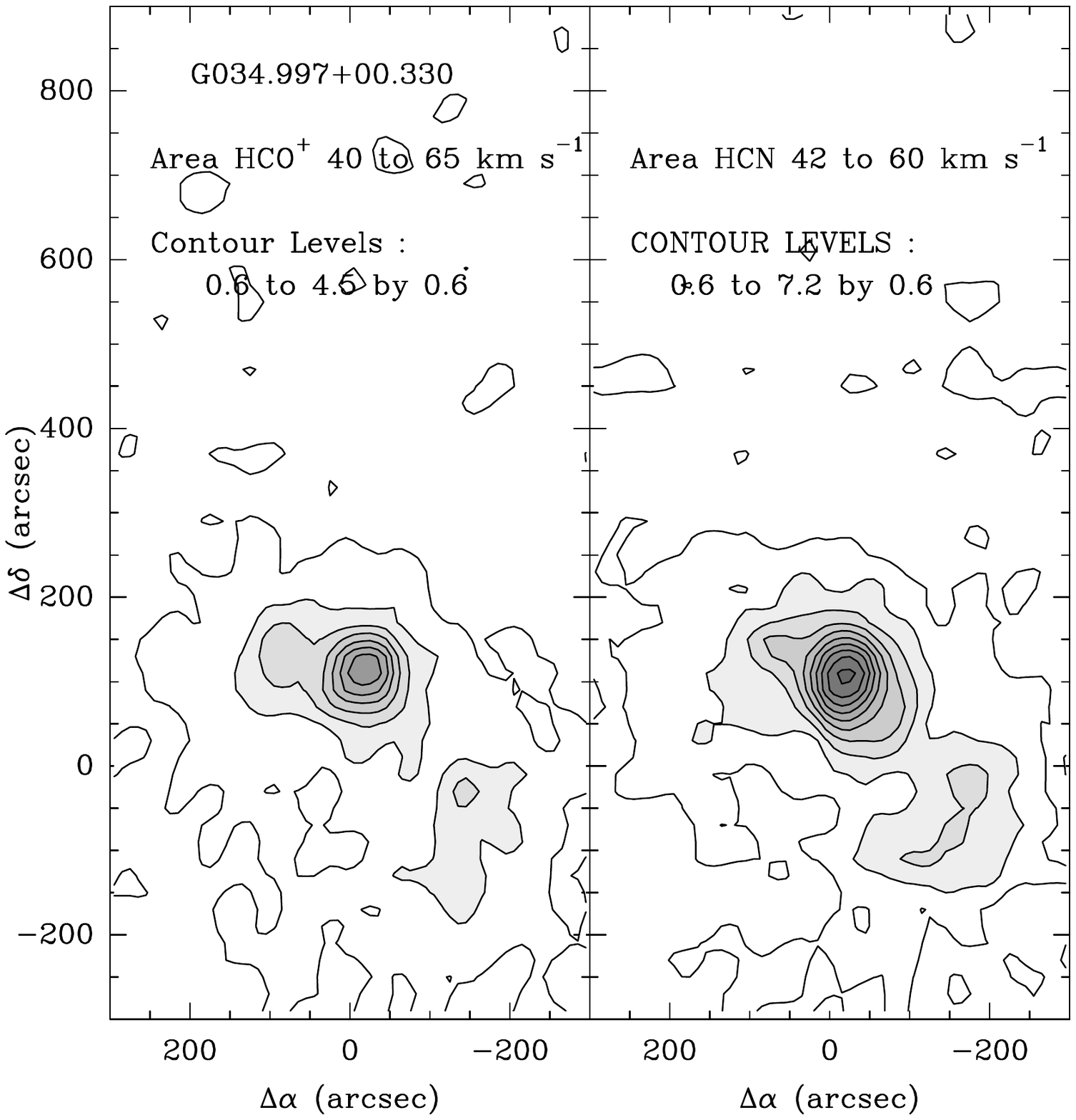}
\caption{The integrated intensities of the \hcop\ and HCN lines
  in G034.997+00.330. The integration is over the range
of 45 to 60 kms\ for \hcop\ and 43 to 65 \kms\ for HCN.
}
\label{G034.997maps}
\end{figure*}

This cloud was mapped in equatorial coordinates with center position
$\alpha_{2000} = 18^h 54^m 01\fs83$, $\delta_{2000} = 01\degr 59\arcmin 18\farcs0$.
The BGPS millimeter continuum emission is shown in figure
\ref{G034.997-2by3}.
The spectra at the peaks indicated in Table \ref{lineprops}
are shown in figure \ref{G034.997spec}.
There is a single velocity component at about 57 \kms.
The HCN line is well fitted with hyperfine components.
The \hcopi\ and \hcni\ lines are about 10\% of the main lines, indicating
only modest optical depth. Nonetheless, we use the \hcopi\ line averaged
over its detected region for the linewidth in computing virial mass, etc.

The contour diagrams of integrated intensity are shown in figure \ref{G034.997maps}.
The BGPS, HCN, and \hcop\ emission regions correspond well. 
However, there is significant
emission at large distances from the peak of the emission. 
The plots of integrated intensity of \hcop\ and HCN  versus distance 
from the peak are shown in figure
\ref{g034ivsr}. The right-most panel shows the smoothed \hcop\ along with 
that from \coo.

\section{\cloudc}


\begin{figure*}[ht]
\centering
\includegraphics[width=0.60\textwidth, angle=0]{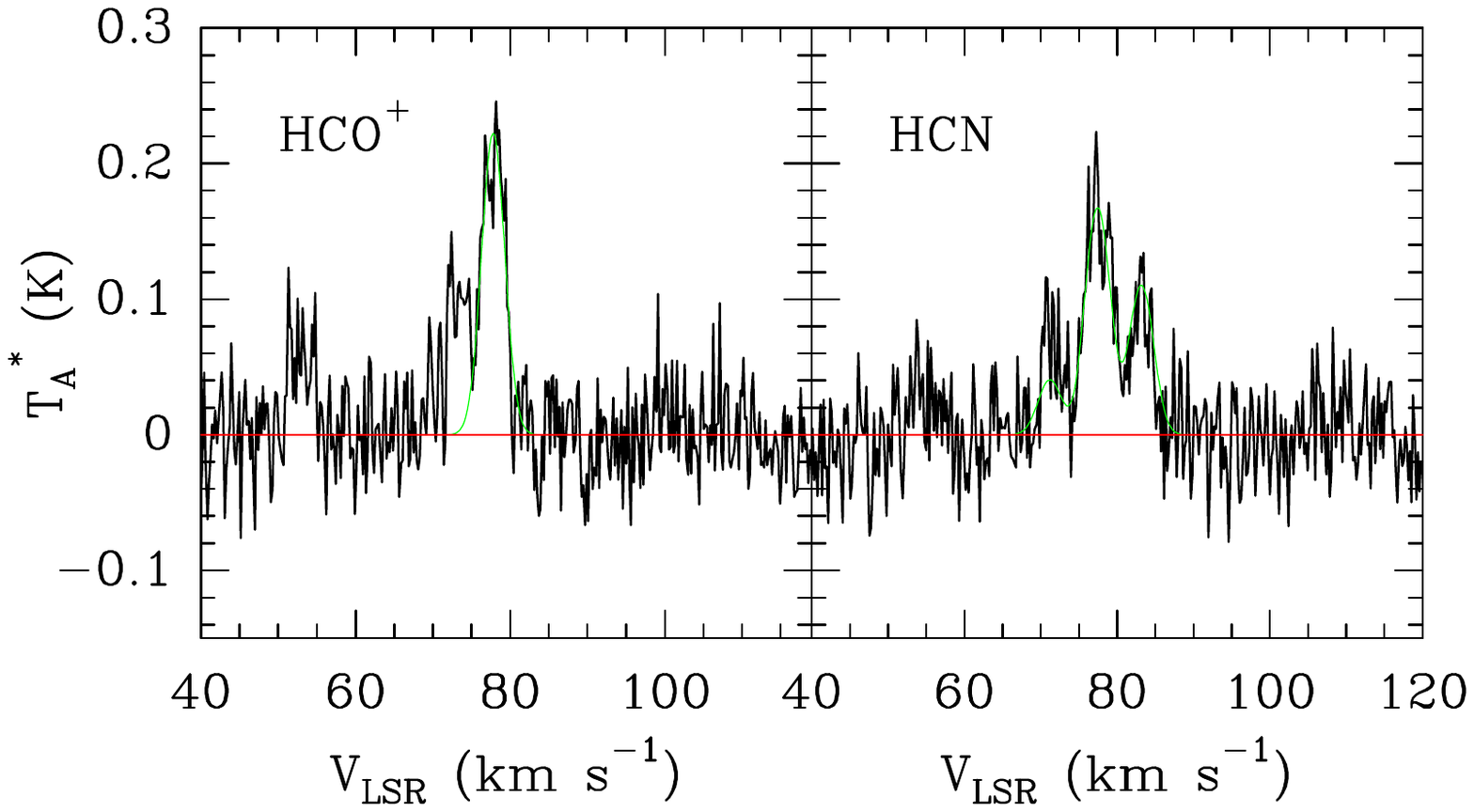}
\caption{The spectra of \hcop and HCN
toward the peak of \cloudc. 
In each plot, the red line shows the zero level and 
the green line shows the fit to the line.
}
\label{G036specs}
\end{figure*}

\begin{figure*}[ht]
\centering
\includegraphics[width=0.6\textwidth]{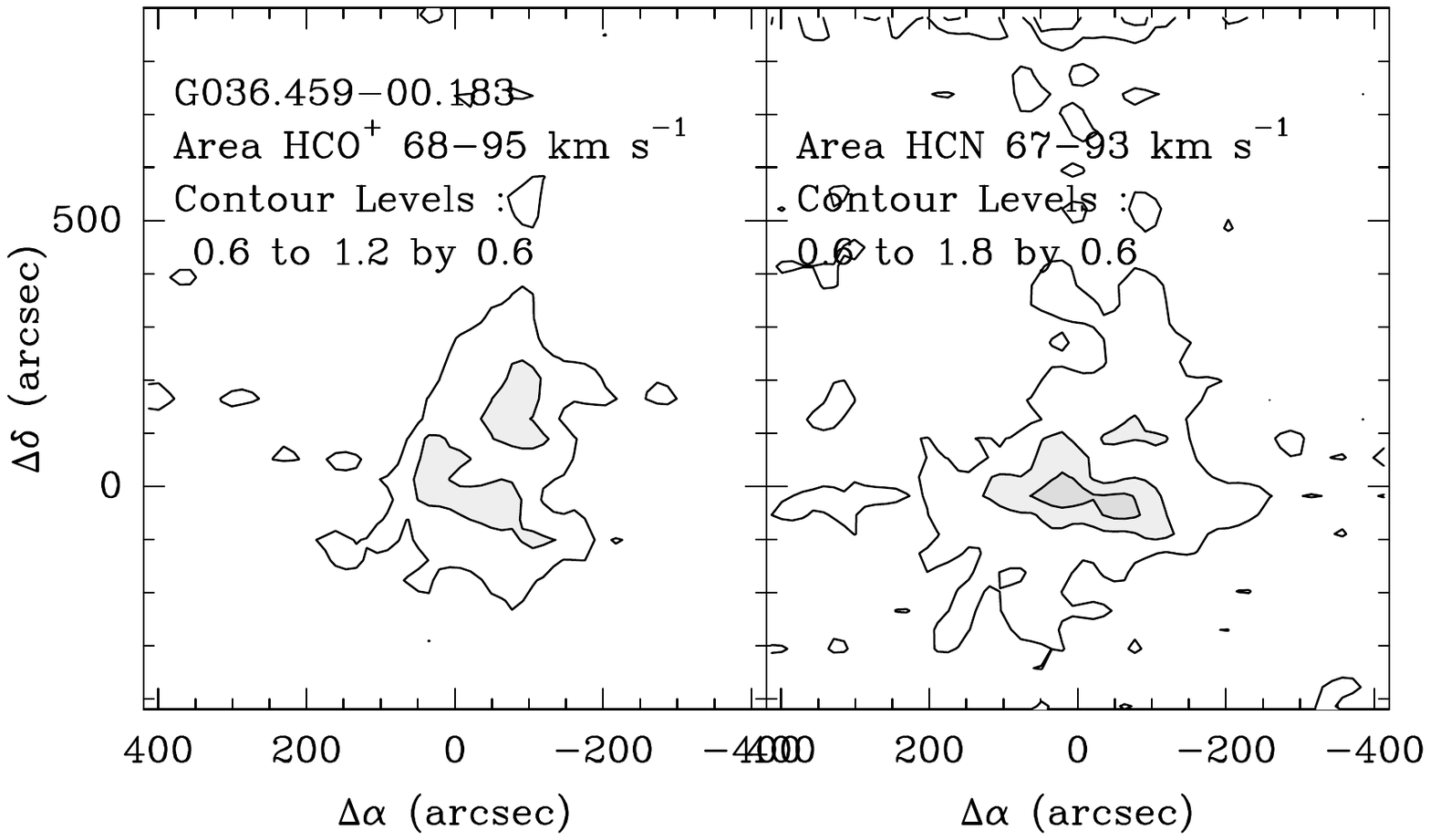}
\caption{The integrated intensities of the \hcop\ and HCN lines
 for the 78 \kms\
component in \cloudc.
}
\label{g036maps}
\end{figure*}

This cloud was mapped in equatorial coordinates with center position
$\alpha_{2000} = 18^h 58^m 31\fs78$, $\delta_{2000} = 03\degr 03\arcmin 18\farcs2$.
The BGPS millimeter continuum emission is shown in figure
\ref{G036.459-2by3}. The spectra are shown in 
 figure \ref{G036specs}.
Neither the \hcopi\ nor the \hcni\ lines were detected at any position.
This cloud
has a primary component at about 73 \kms,which is associated with the \hii\ 
region, based on the recombination line velocity. 
A secondary component at about 55 \kms\ appears in the spectrum
but it is not related to this cloud, so we do not analyze it. 
The two components nearly overlap, but are separable at about
67 to 68 \kms. The hyperfine structure was used to fit the
HCN lines, but the integrated intensity was integrated over all hyperfine
components.

The  maps of \hcop\ and HCN are shown in figure \ref{g036maps}.
The \hcop\ and HCN integrated intensity generally map similarly, though
there are clearly differences in detail.  The emission is not well peaked, and 
the half-power contour is nearly as large as the region of detected 
emission, so the properties are poorly defined.

\section{\cloudd}


\begin{figure*}[ht]
\centering
\includegraphics[width=0.60\textwidth, angle=0]{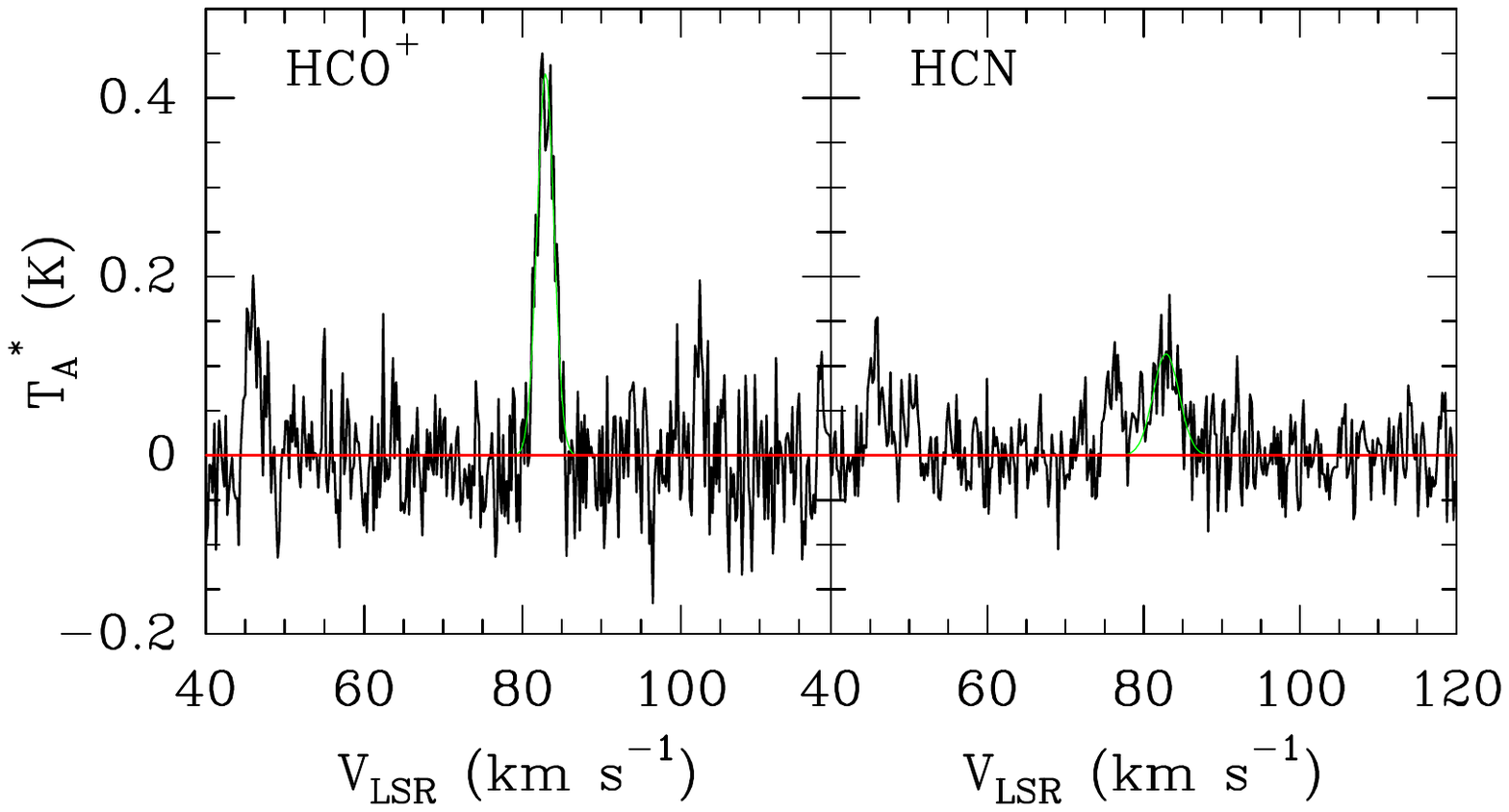}
\caption{The spectra of \hcop and HCN
toward their peaks in \cloudd. 
The \hcop\ line is at position $(-100, -160)$; the HCN line
is an average of the 9 positions surrounding offset $(-40, -120)$.
The HCN line was not well fitted with
hyperfine structure so only the main peak was fitted. Note that the peak
positions are somewhat different for the two species.
In each plot, the red line shows the zero level and 
the green line shows the fit to the line.
}
\label{G037spec}
\end{figure*}

\begin{figure*}[ht]
\centering
\includegraphics[width=0.40\textwidth]{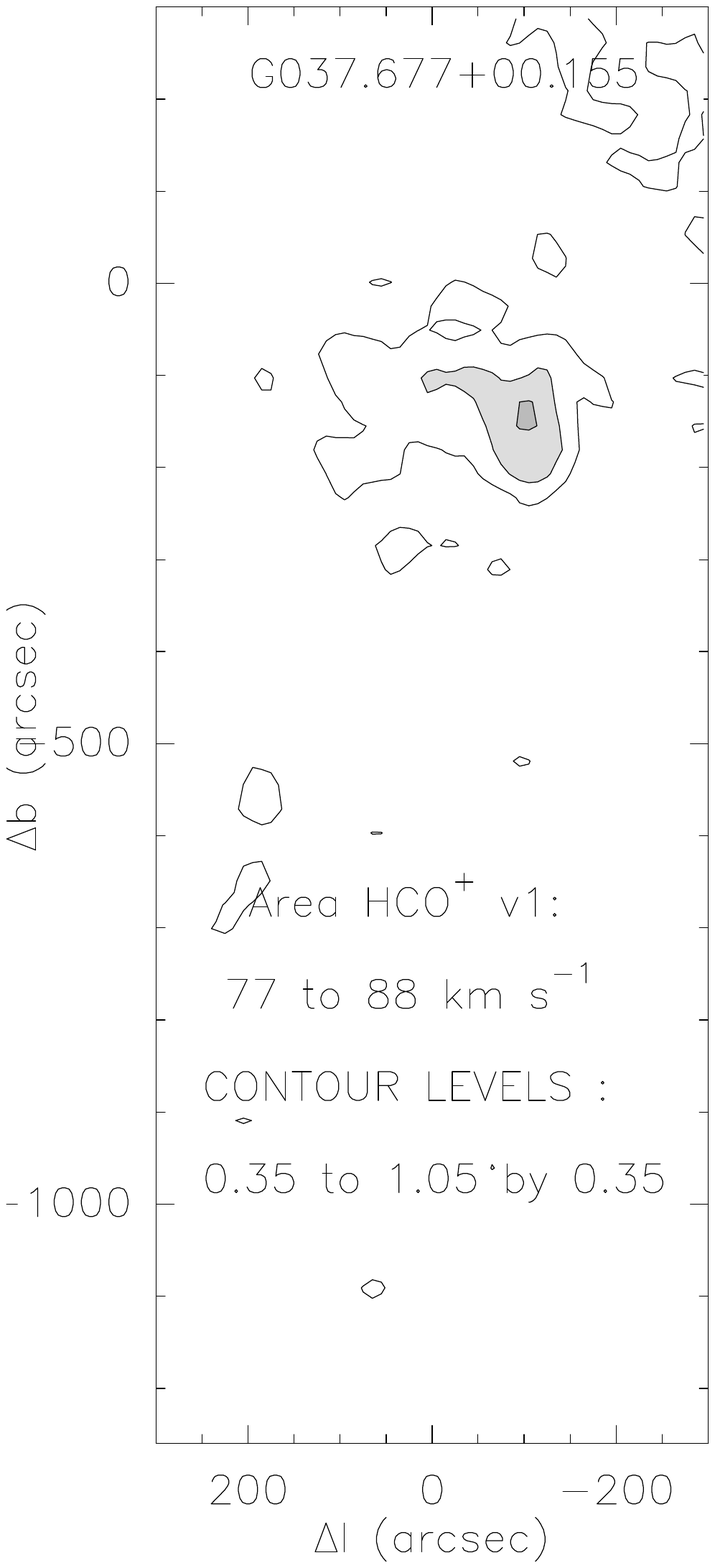}
\includegraphics[width=0.40\textwidth]{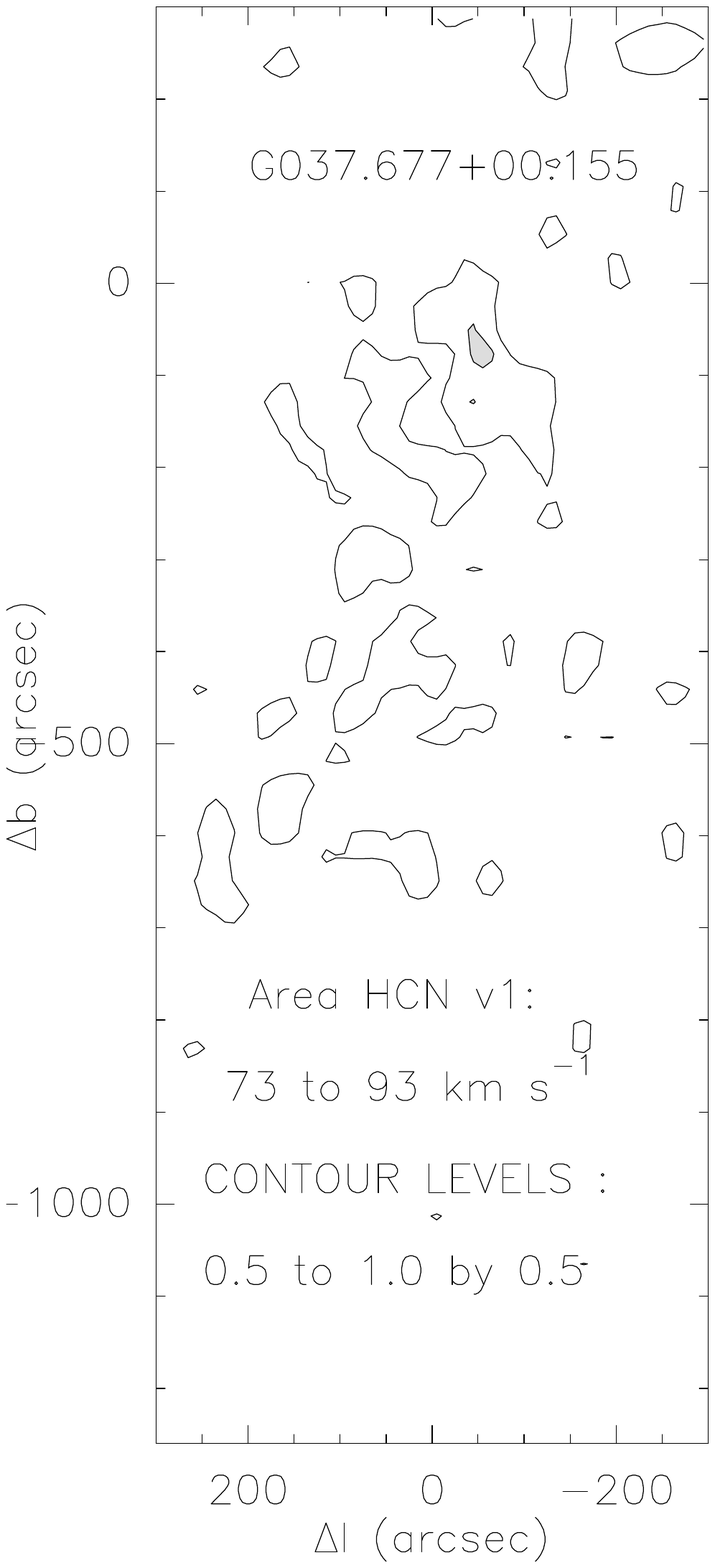}
\caption{The integrated intensities of the \hcop\ and HCN lines
  in G037.677+00.155  are shown for the velocity component near
$82$ \kms. 
}
\label{G037maps}
\end{figure*}

This source was mapped in galactic coordinates centered on the
source name position, $l = 37.677$, $b = 0.155$.
The BGPS millimeter continuum emission is shown in figure
\ref{G037.677-2by3}. There are at least two
velocity components, one near 45 \kms\ and one near 83 \kms.
The second one is near the velocity of the radio recombination line
 for this HII region,
so we focus on that. Figure \ref{G037spec} shows the spectra.
Nine positions around
a nominal peak were averaged for HCN to produce the spectrum.
 The data were not well fitted with hyperfine structure,
so we isolated the velocities around 83 \kms\ to be fitted. 

The \hcop\ and HCN integrated intensity maps  (figure \ref{G037maps})
show extended weak emission so the
peaks are not well defined. The peaks of the two lines differ.
 The weak emission is surprising because the cloud is relatively
massive and has a substantial star formation rate and mass of dense gas, based
on the submillimeter continuum data. However, the cloud was difficult
to define as it exists in a region of considerable line confusion in
\coo.

\section{\cloude}


\begin{figure*}[ht]
\centering
\includegraphics[width=0.60\textwidth, angle=0]{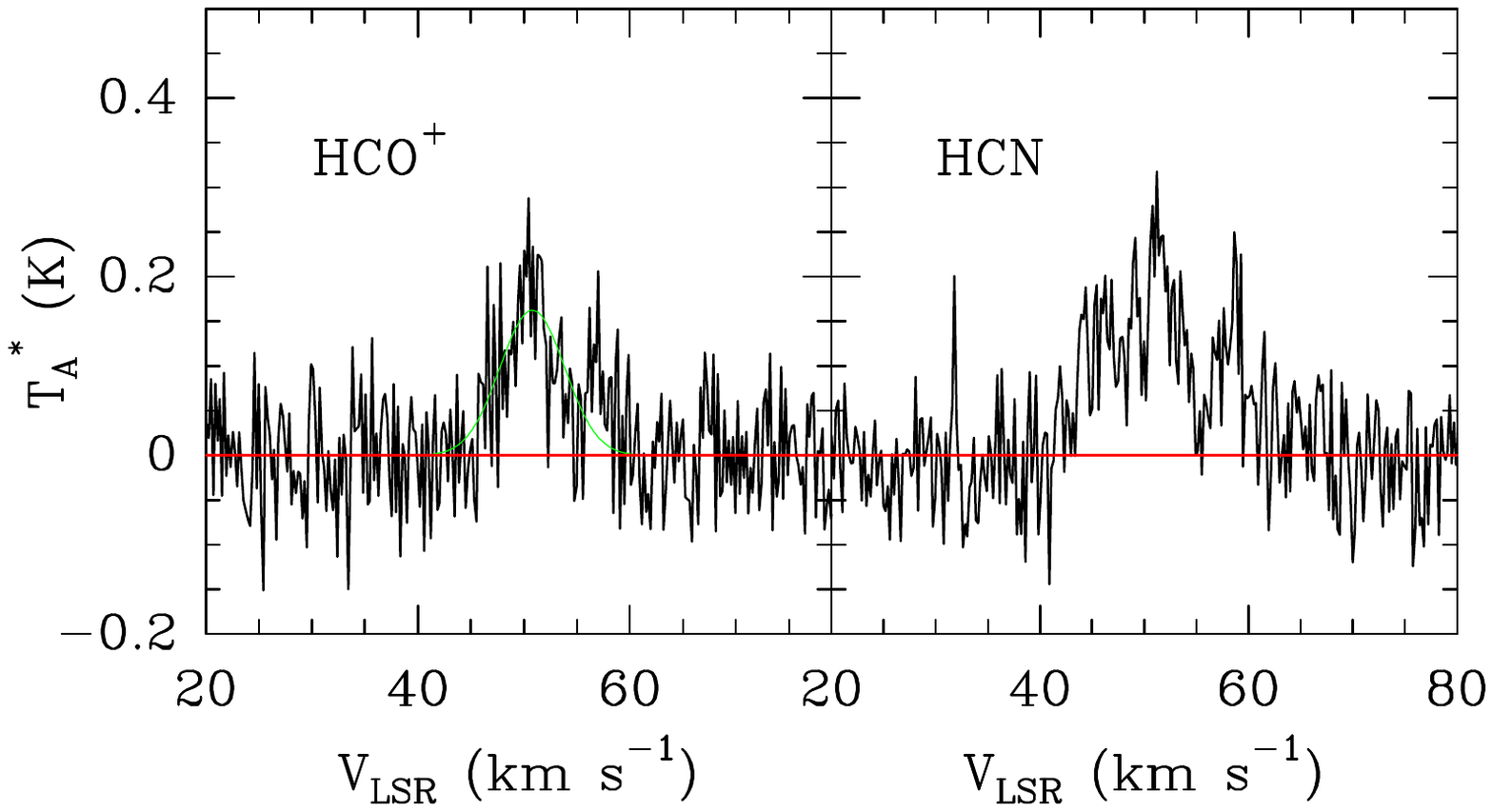}
\caption{The spectra of \hcop\ and HCN
toward their peaks in \cloude. 
The \hcop\ line is is an average of the 9 positions surrounding offset
 $(-420, 60)$; the HCN line
is an average of the 9 positions surrounding offset $(-400, 80)$.
The HCN line was not well fitted with
hyperfine structure. Note that the peak
positions are somewhat different for the two species.
In each plot, the red line shows the zero level and 
the green line shows the fit to the line.
}
\label{G045spec}
\end{figure*}

\begin{figure*}[ht]
\centering
\includegraphics[width=0.50\textwidth, angle=0]{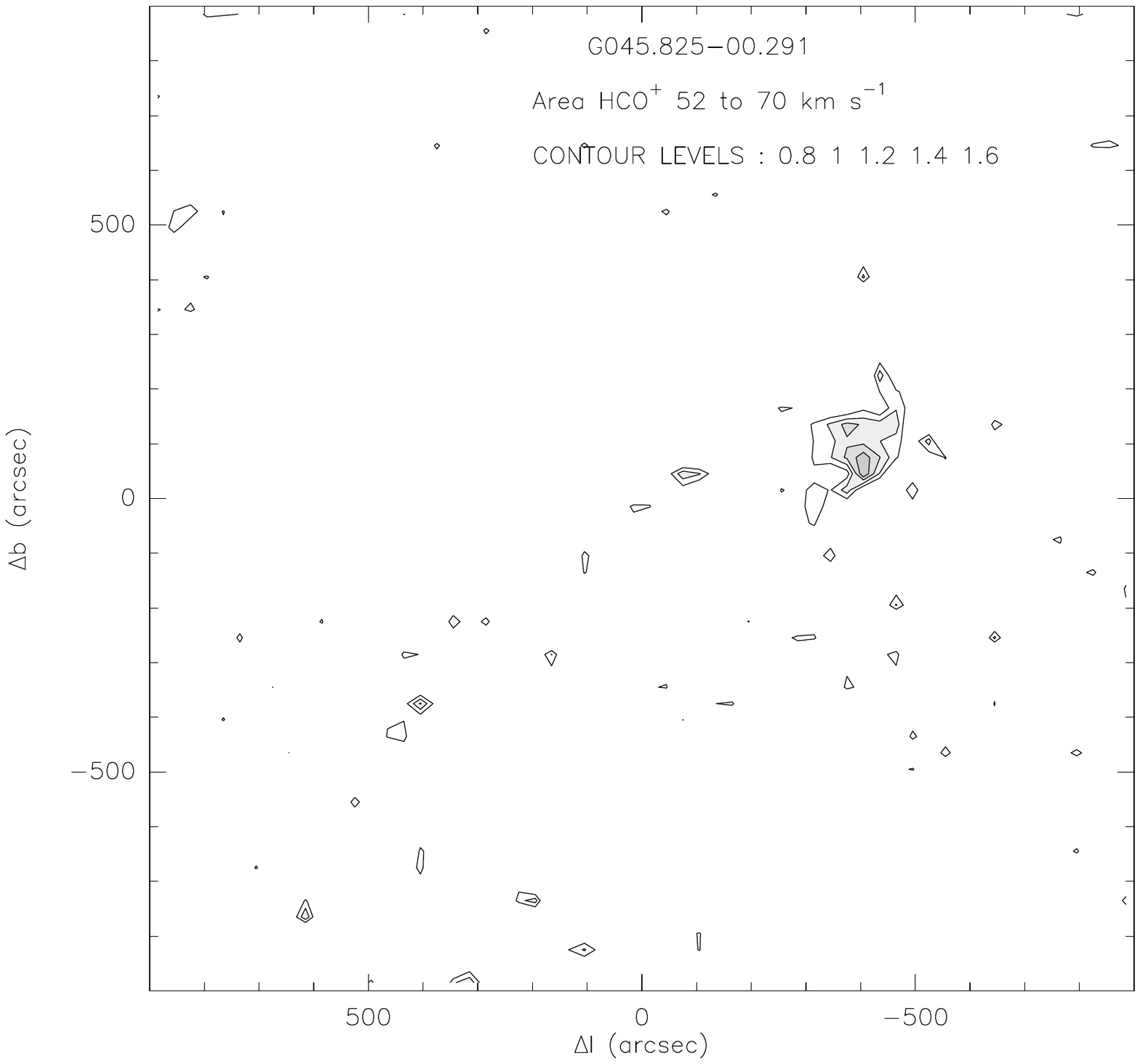}
\includegraphics[width=0.50\textwidth, angle=0]{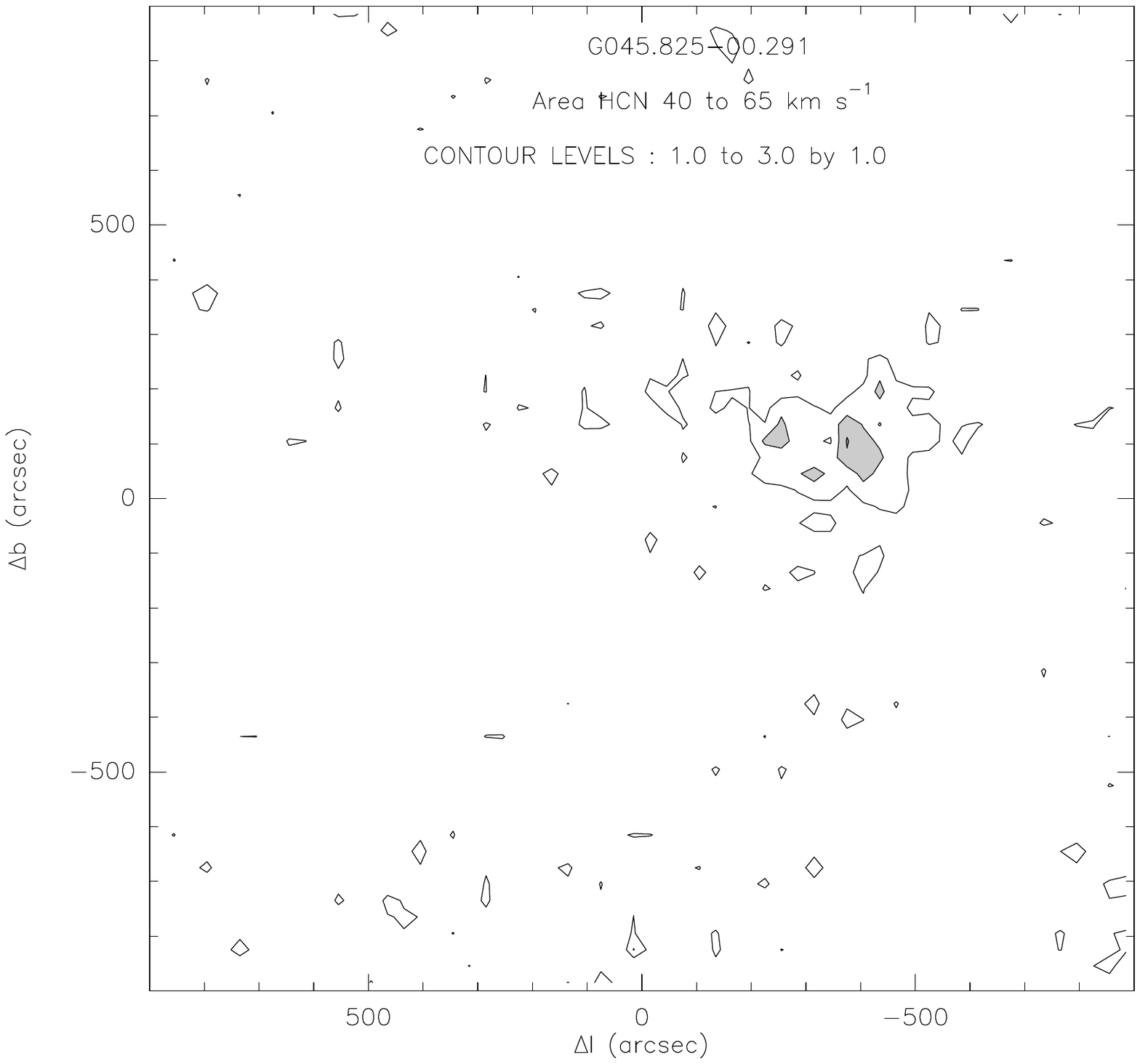}
\caption{The integrated intensities of the \hcop\ and HCN lines
 for both components in \cloude  are shown.
}
\label{g045maps}
\end{figure*}

This source was mapped in galactic coordinates centered on the
source name position, $l = 45.825$, $b = -0.291$
The BGPS millimeter continuum emission is shown in figure
\ref{G045.825-2by3}.
 There are several
peaks in the BGPS data. Only the ``northwest'' one of these shows 
up in the \hcop/HCN maps (figure \ref{g045maps}).

This source has two velocity components that are barely separable
in \hcop\ and difficult to separate in HCN. Both \hcop\ and HCN 
lines are weak, but
HCN is a bit stronger in integrated intensity. Figure \ref{G045spec}
shows the spectra. We average over the 9 positions around the peak
to improve the noise since the lines are very similar.
Since  both velocity features peak
in the same area, we include both in the maps of intensity. 
The isotopologue lines were not detected and did not
provide useful constraints on optical depth. We plot the contours of
emission for both main species (Fig. \ref{g045maps}); the peaks are
slightly different for the two species, but generally consistent.

For \hcop\ and HCN, we use the velocity width and area of the 50.5 \kms\
feature to compute virial mass, etc. However, the HCN is confused by the
hyperfine structure.

\section{G046.495-00.241}


\begin{figure*}[ht]
\centering
\includegraphics[width=0.60\textwidth, angle=0]{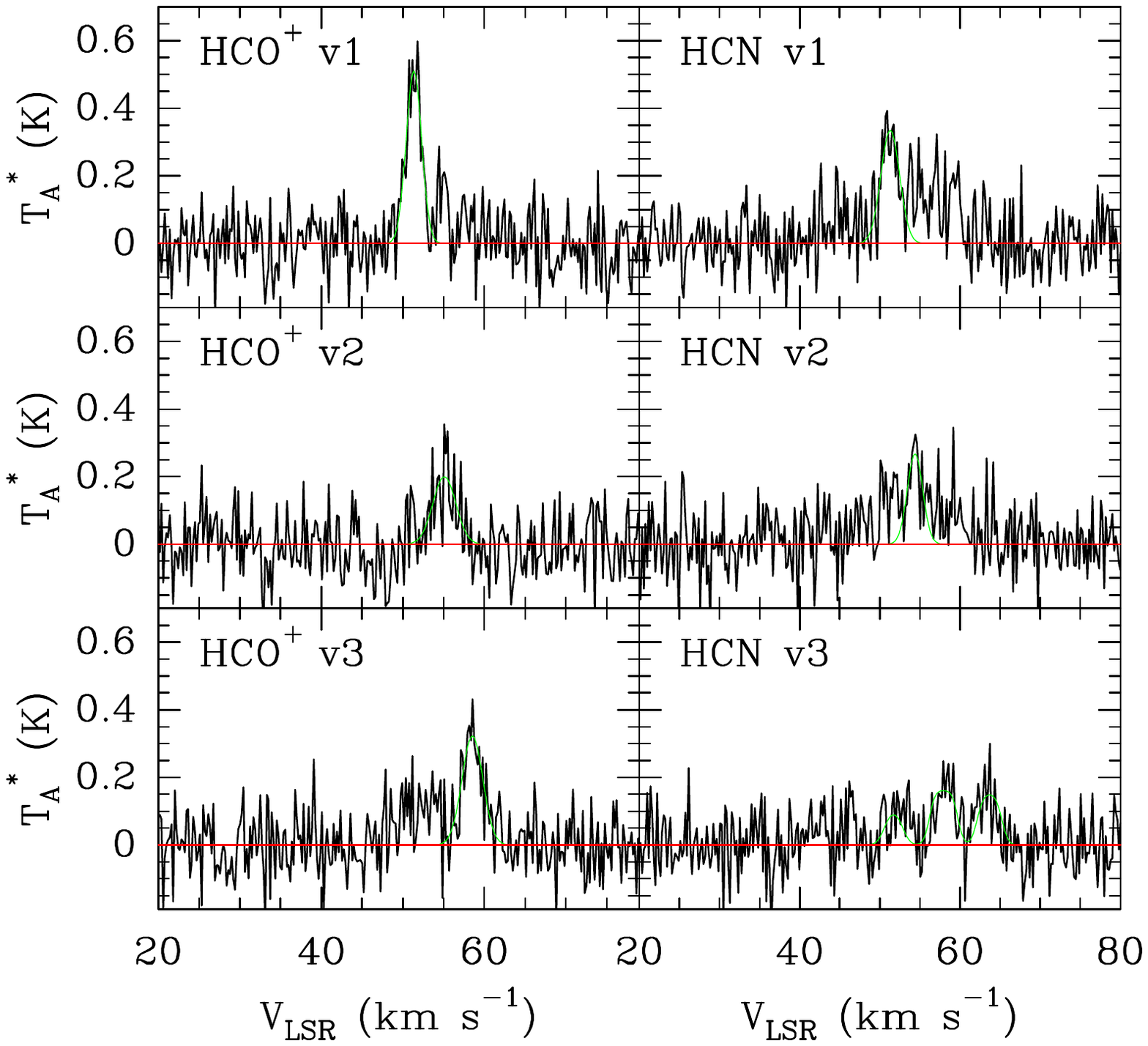}
\caption{The spectra of \hcop\ and HCN
toward the peaks for each velocity component in \cloudf. 
The positions are those nearest the peak positions listed in Table
\ref{lineprops}.
The HCN lines for the first two velocity components were not well fitted with
hyperfine structure so a single Gaussian was fitted to the
main peak. 
In each plot, the red line shows the zero level and 
the green line shows the fit to the line.
}
\label{G046spec}
\end{figure*}

\begin{figure*}[ht]
\centering
\includegraphics[width=0.50\textwidth]{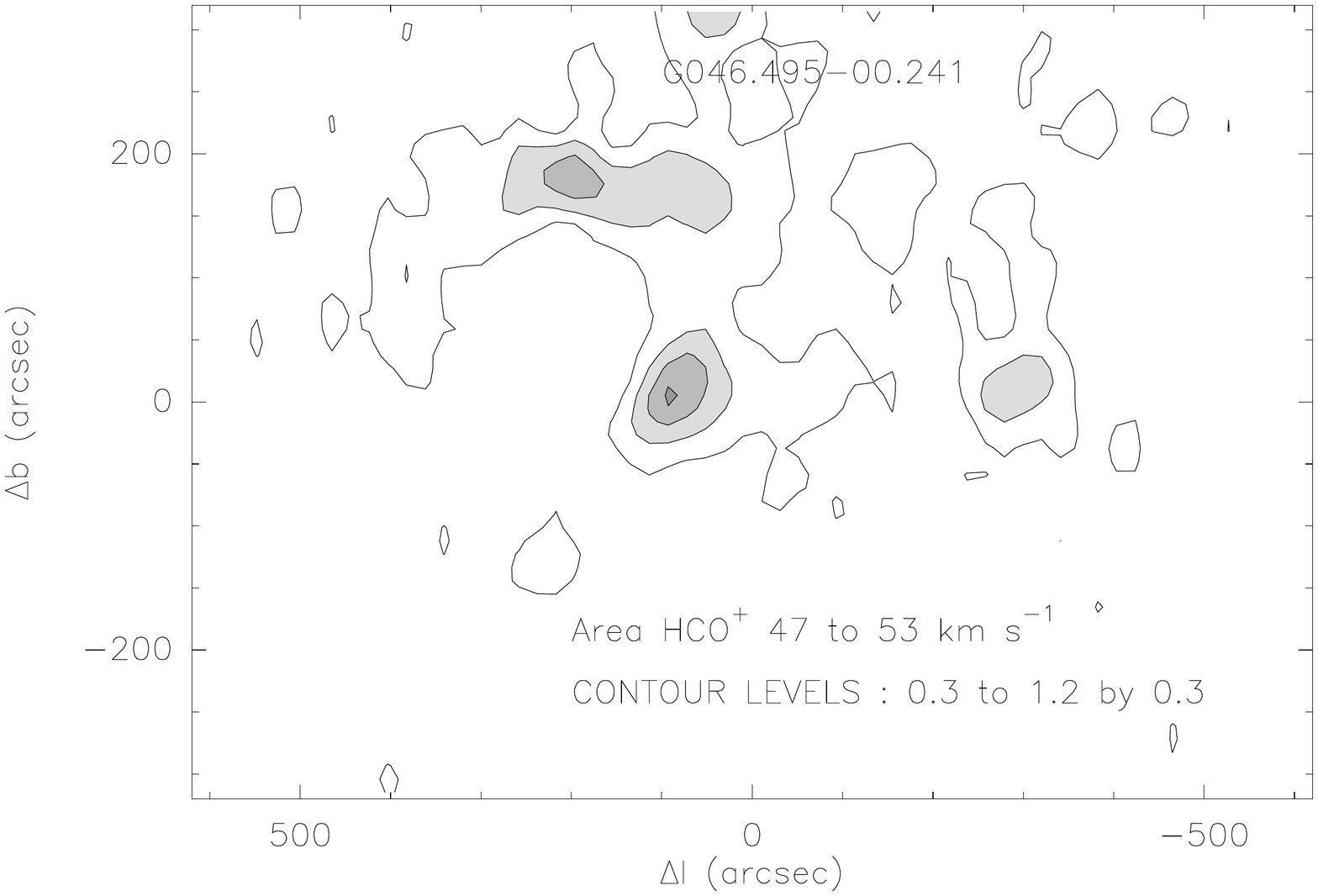}
\includegraphics[width=0.50\textwidth]{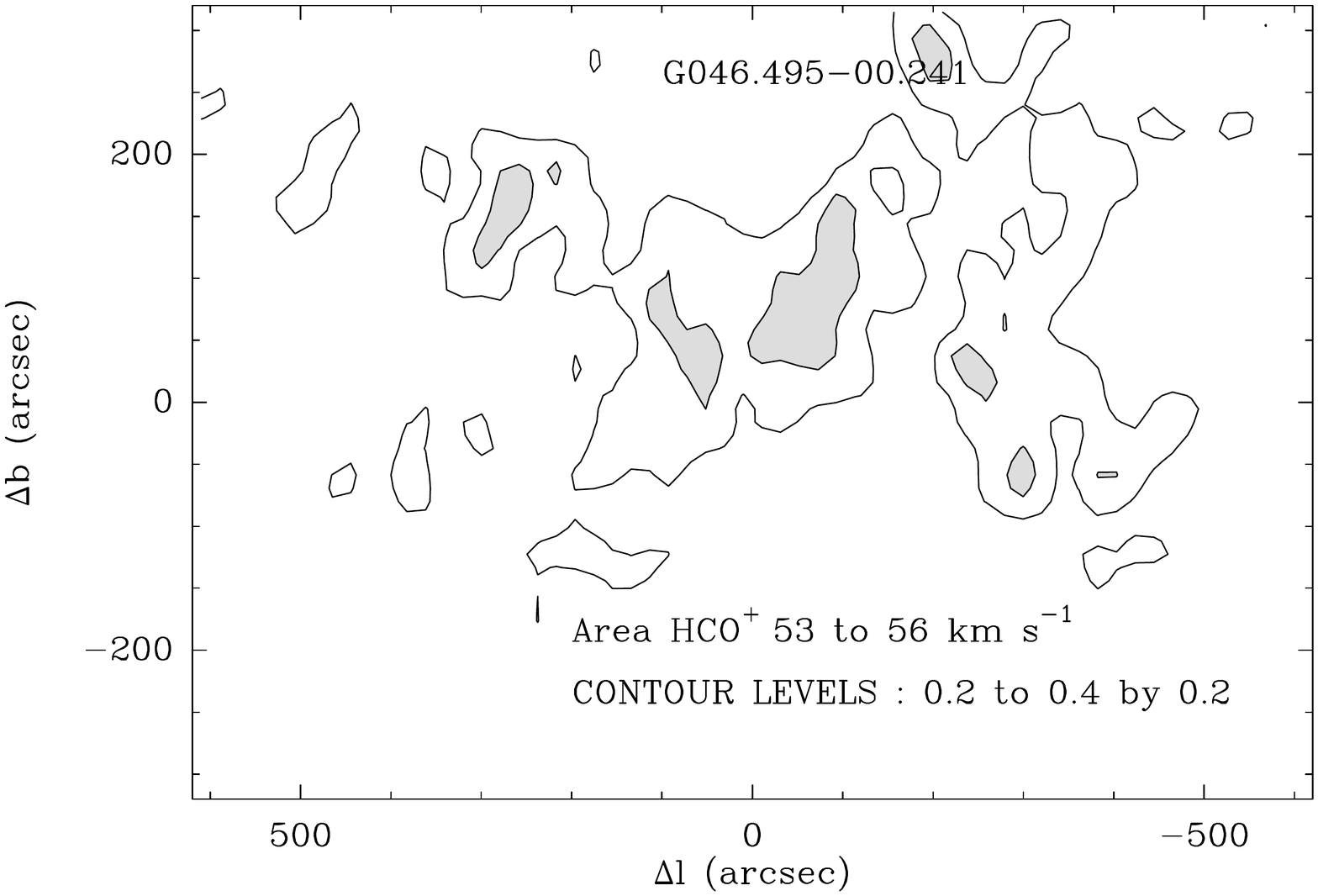}
\includegraphics[width=0.50\textwidth]{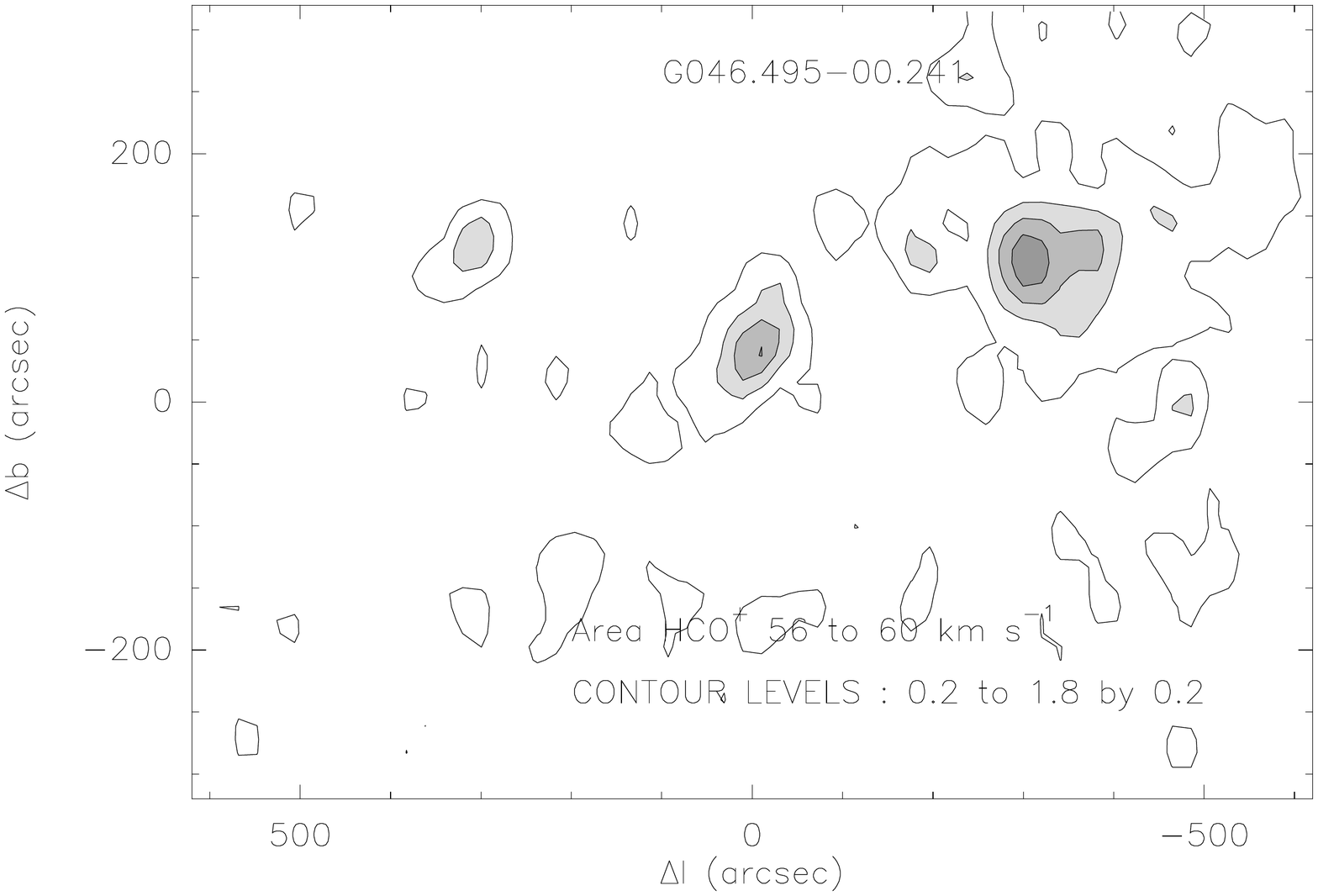}
\caption{The integrated intensities of the \hcop\  lines
  in G046.495-00.241  are shown for the three different velocity 
components.}
\label{G046mapsvx}
\end{figure*}


\begin{figure*}[ht]
\centering
\includegraphics[width=0.50\textwidth]{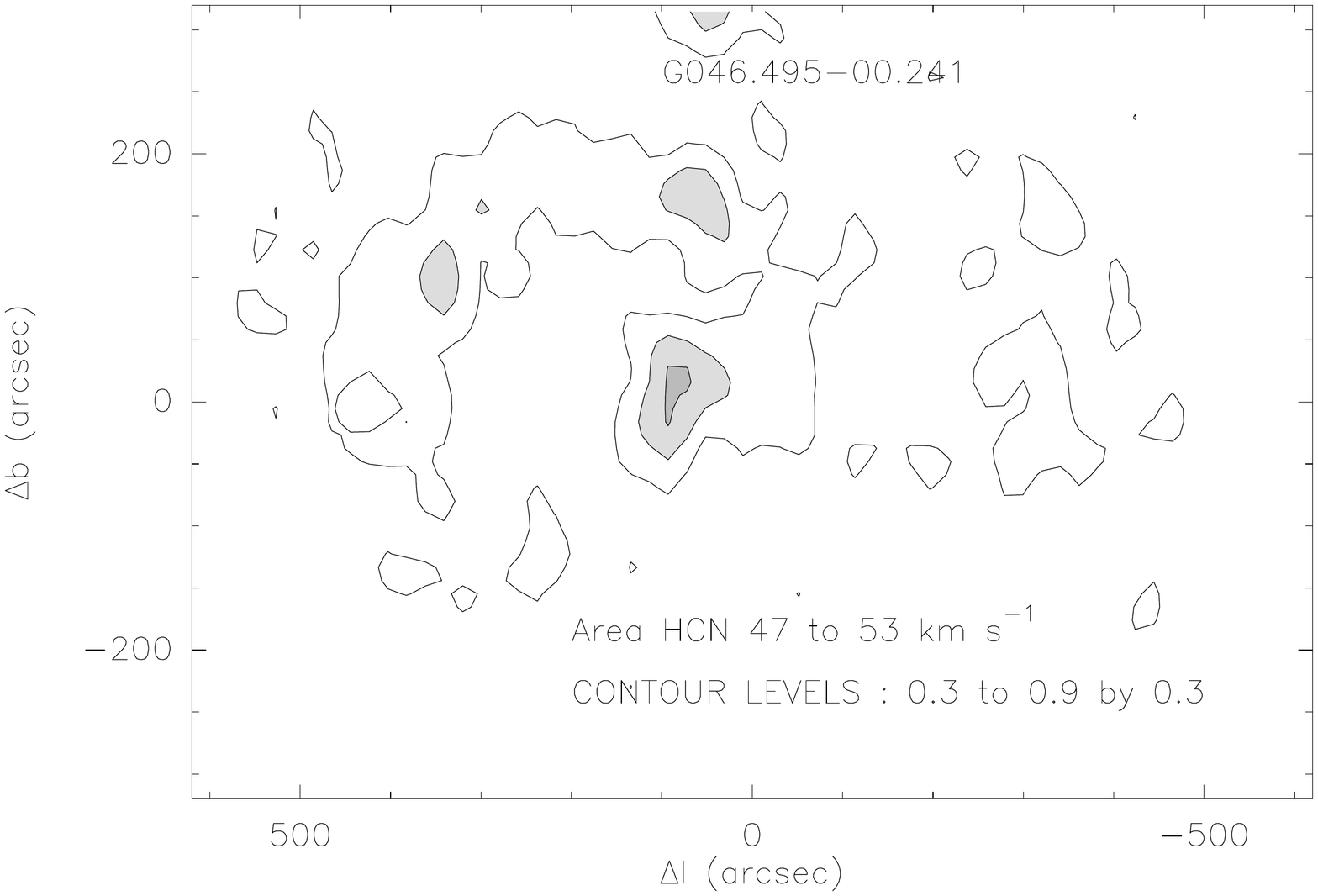}
\includegraphics[width=0.50\textwidth]{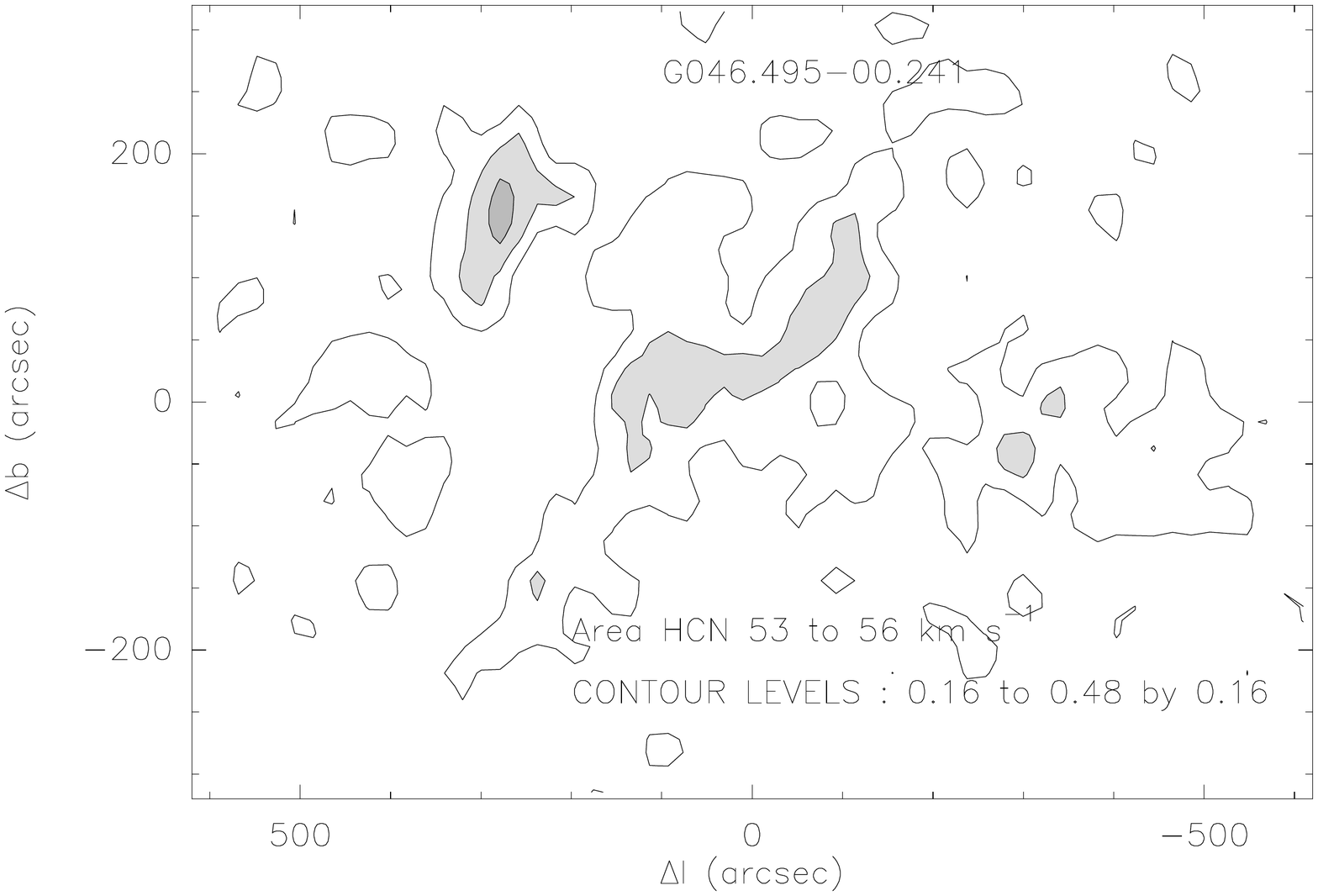}
\includegraphics[width=0.50\textwidth, angle=-90]{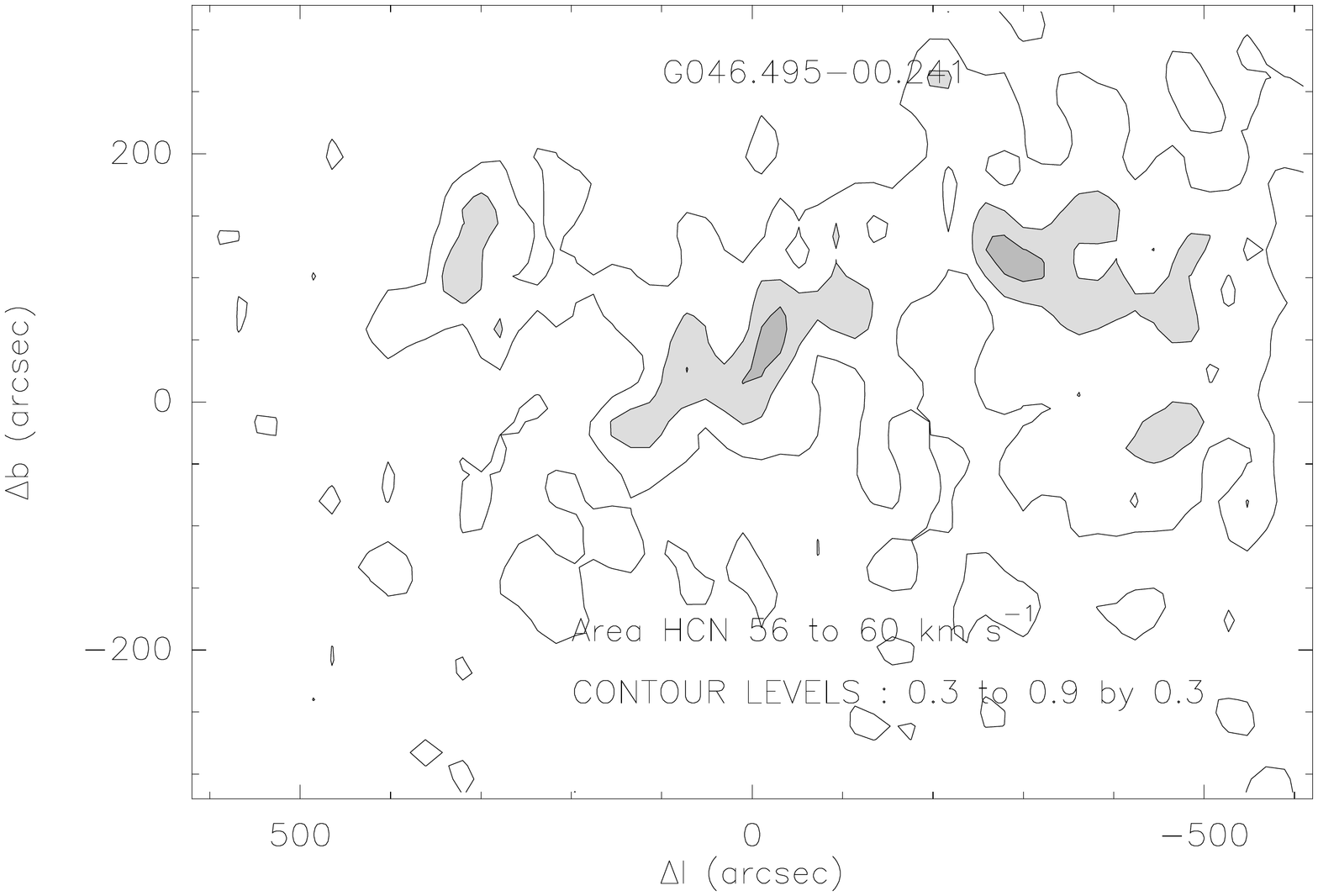}
\caption{The integrated intensities of the HCN lines
  in G046.495-00.241  are shown for the three different velocity 
components.}
\label{G046mapshcnvx}
\end{figure*}

This source was mapped in galactic coordinates centered, not on
the source name, but on  $l = 46.400$, $b = -0.241$, 
because the source name was that of a \mm\
continuum source in the eastern part of the cloud.
The BGPS millimeter continuum emission is shown in figure
\ref{G046.495-2by3}.
There are three regions
of continuum emission, which we refer to as west, central, and east
in order of increasing longitude.
Both \hcop\ and HCN appear to 
have several velocity components (Fig. \ref{G046spec})
and three separated emission regions when the emission
is integrated over all velocities, which
correspond roughly to the continuum peaks. 
There were no detections of the $^{13}$C isotopologues, so 
we concluded that the velocity structure was unlikely to
be caused by self-absorption.

It was possible to  separate the three velocity components in
\hcop\ and to make contour maps.
The maps of integrated intensity are shown in figure \ref{G046mapsvx}.
The spectra at the peaks of each component are shown in
figure \ref{G046spec}.
The lowest velocity component, v1, peaks most strongly on the central 
peak, but has secondary peaks to the north and west that do not
correspond exactly to the peaks of the other components, but overlap
with them. 
The middle velocity component (v2) emits over an extended region, but
weakly, with no strong peaks. We picked the most central and largest
in size peak, which is south and east of the peak of v1.
The highest velocity component denoted v3, peaks most strongly on 
the western (lower $l$) peak, but has secondary maxima near the other peaks.
The other lines are largely absent from the western peak. Because
the individual velocity components are narrow, the rms in the integrated
intensity is small, so it is possible to draw more contours.
The secondary peaks and extended plateaus are reflected in the secondary
maxima, but the main peak is reasonable well defined.
The separation for HCN is more difficult because of the hyperfine
structure, so we used the velocity intervals from the \hcop\ analysis
for components v1 and v2. The hyperfine structure was more visible and
separate for v3, so we fit that component. 
The separation into velocity components would not be possible in
observations of other galaxies.
The contour maps 
for each species and component in
(fig. \ref{G046mapshcnvx})
for each component indicate
the complexity of overlapping regions.

\clearpage

\bibliographystyle{aasjournal}
\bibliography{cite,trao_mh} 

\end{document}